\documentclass[]{aastex631}

\usepackage[normalem]{ulem}

\newcommand{\co}{\mbox{${\rm C/O}$}}

\newcommand{\ominusc}{\mbox{$\mathrm{O}-\mathrm{C}$}}
\newcommand{\Msun}{\mbox{$\mathrm{M}_{\odot}$}}
\newcommand{\Lsun}{\mbox{$\mathrm{L}_{\odot}$}}

\newcommand{\cm}{\textup{ cm}}

\newcommand{\s}{\textup{ s}}
\newcommand{\K}{\textup{ K}}
\newcommand{\R}{\textup{ R}}

\newcommand{\HH}{\textup{H}}

\newcommand{\He}{\textup{He}}
\newcommand{\e}{\textup{e}^-}

\newcommand{\g}{\textup{ g}}

\newcommand{\kb}{k_{\rm B}}
\newcommand{\rmd}{\,{\rm d}}
\newcommand{\kR}{\mbox{$\kappa_{\rm R}$}}
\newcommand{\Zsun}{\mbox{$\mathrm{Z}_{\odot}$}}

\shorttitle{Low-Temperature Gas Opacities with \AE SOPUS\,2.0}
\shortauthors{Marigo et al.}
\usepackage{multirow}
\usepackage{booktabs}
\usepackage{threeparttable}
\usepackage{courier}

\begin{document}
\title{Updated Low-Temperature Gas Opacities with \AE SOPUS\,2.0}

\correspondingauthor{Paola Marigo}
\email{paola.marigo@unipd.it}

\author[0000-0002-9137-0773]{Paola Marigo}
\affiliation{Department of Physics and Astronomy G. Galilei,
University of Padova, Vicolo dell'Osservatorio 3, I-35122, Padova, Italy}
\author[0000-0001-9848-5410]{Bernhard Aringer}
\affiliation{Department of Astrophysics,  University of Vienna, T\"urkenschanzstrasse 17, 1180 Vienna, Austria}
\author[0000-0002-6301-3269]{L\'eo Girardi}
\affiliation{INAF-Osservatorio Astronomico di Padova  Vicolo dell’Osservatorio 5, I-35122 Padova, Italy}
\author[0000-0002-5434-1973]{Alessandro Bressan}
\affiliation{SISSA, via Bonomea 265, I-34136 Trieste, Italy}



\begin{abstract}
This work introduces new low-temperature gas  opacities, in the range $3.2 \la \log (T/{\rm K}) \la 4.5$, computed with the  \texttt{\AE SOPUS} code under the assumption of thermodynamic equilibrium \citep{Marigo_Aringer_2009}. In comparison to the previous version \texttt{\AE SOPUS\,1.0}, we updated and expanded molecular absorption to include 80 species, mostly using the recommended line lists currently available from the \texttt{ExoMol} and \texttt{HITRAN} databases. Furthermore, in light of a recent study, we revised the H$^-$ photodetachment cross section,  added the free-free absorption of other negative ions of atoms and molecules, and updated the collision-induced absorption due to H$_2$/H$_2$, H$_2$/H, H$_2$/He, and H/He pairs. Using the new input physics, we computed tables of Rosseland mean opacities for several scaled-solar chemical compositions, including \citet{Magg_etal_22}'s most recent one, as well as $\alpha$-enhanced mixtures. The differences in opacity between the new \texttt{\AE SOPUS\,2.0} and the original \texttt{\AE SOPUS\,1.0} versions, as well as other sets of calculations, are discussed. The new opacities are released to the community via a dedicated web-page that includes both pre-computed tables for widely used chemical compositions, and a web-interface for calculating opacities on-the-fly for any abundance distribution. 
\end{abstract}

\keywords{Stellar atmospheric opacity(1585) --- Astrochemistry(75) --- Molecular physics(2058) --- Chemical abundances(224)}


\section{Introduction} \label{sec:intro}
Low-temperature gas opacities, in the approximate temperature range, $1500 \la T/{\rm K} \la 10000$, play a pivotal role in a variety of astrophysical applications.
The continuous absorption of the negative hydrogen ion H$^-$, for example, is one of the most important opacity sources in our Sun's atmosphere, the strength of which is also dependent on the availability of free electrons provided by elements with relatively low ionization potentials, primarily Mg, Si, Fe, Al, C, Ca.
Molecular opacities dominate the atmospheres of cool stars with temperatures $T<4500$ K (pre-main sequence stars, main-sequence red dwarfs, red giants and supergiants). Following the discovery of thousands of exoplanets by space missions such as Kepler and Corot, the demand for molecular opacities has grown tremendously in the last decades \citep{EXOMOL_2012MNRAS.425...21T, Grimm_etal_21, Chubb_etal_21}.
Furthermore, the advent of high-resolution and large spectroscopic surveys has revealed a wide range of chemical patterns at the photospheres of stars \citep[see][for a thorough review]{Jofre_etal_19}, which must be addressed properly by theory with stellar models that include consistent chemical composition and opacity.

For many years the Wichita State University group \citep[e.g.,][]{AlexanderFerguson_94, Ferguson_etal_05} has been the historical supplier of low-temperature opacities. Typically, these authors provide opacity tables for scaled-solar or $\alpha$-enhanced mixtures,  primarily designed for stellar structure computations.
Other groups have produced low-temperature opacities,  suitable for brown dwarfs and planetary atmospheres \citep{SharpBurrows_07, Grimm_etal_21, Chubb_etal_21}, for protoplanetary disks \citep{Semenov_etal_03}, for primordial matter \citep{Harris_etal_04, Mayer_Duschl_05}, 
for carbon- and nitrogen-enriched asymptotic giant branch stars \citep{LedererAringer_09}.

In 2009 \citeauthor{Marigo_Aringer_2009} developed the \texttt{\AE SOPUS} code, which  solves the equation of state for over 800 chemical species (300 atoms/ions and 500 molecules)  and calculates the Rosseland mean opacities for any combination of abundances  assuming thermodynamic equilibrium.
The primary goal of that work was to greatly expand public access to Rosseland mean opacity data in the low-temperature regime.
We created a web-interface (\href{http://stev.oapd.inaf.it/aesopus}{http://stev.oapd.inaf.it/aesopus}) that allows users to compute and quickly retrieve Rosseland mean opacity tables based on their specific needs, with complete control over the chemical composition of the gas  (individual abundances can be set for $92$ atomic species, ranging from Hydrogen to Uranium).

A distinguishing quality of \texttt{\AE SOPUS} is its quick performance, which is made possible by an optimized application of the opacity sampling method.
The typical computation time for one table at fixed chemical composition, arranged with the default parameter grid (temperature $T$ and $R=~\rho/(T/10^6\,{\rm K})^3$), i.e. containing ${\rm N}_T \times {\rm N}_R = 67 \times 19 = 1273$ opacity values, is less than 50 s with a 2.0 GHz processor. Thanks to this property, 
we could make \texttt{\AE SOPUS} available online through a web-interface that generates opacity tables in real time, with maximum flexibility and minimal computational cost.
Furthermore, one of the interface's most useful features is the ability to generate a large database of opacity tables with varying C, N, and O abundances.
This is critical for accurately modeling the atmospheric layers of asymptotic giant branch (AGB) stars, the surface composition of which is frequently altered by mixing episodes (third dredge-up) and nucleosynthesis in the convective envelope (hot-bottom burning), as well as massive and super-giant stars.

The \texttt{\AE SOPUS} tool has being used by several  groups to model, e.g.,  AGB stars \citep{Ventura_etal_2018,Karakas_Lugaro_2016}, super-AGB stars \citep{Gil-Pons_etal_2018}, long period variables \citep{Trabucchi_etal_2021}, Supernova light curves \citep{Takei_Shigeyama2020}, white dwarfs \citep{Althaus_etal_2010}. An extended grid of \texttt{\AE SOPUS} tables with varying CNO abundances is currently implemented in the \texttt{MESA} code \citep{Paxton_etal_11}.

In this paper, we present \texttt{AESOPUS\,2.0}, a renewed version of our chemistry and opacity code. We have significantly updated and expanded its ingredients, primarily in relation to the opacity sources, partition functions, and computation-speeding strategies. In addition, we have included the most recent solar chemical compositions published in literature \citep[e.g.,][]{Magg_etal_22, Asplund_etal_21}.
We generated a large number of Rosseland mean opacity tables for several values of metallicity and hydrogen abundance, solar compositions, and $\alpha$-enhanced mixtures. They can be found at the URL \url{http://stev.oapd.inaf.it/aesopus_2.0/tables}. The \texttt{\AE SOPUS} web-interface for Rosseland mean opacities on demand  has also been refurbished.

The paper is structured as follows. Section~\ref{sec_aesopus} recaps the main \texttt{\AE SOPUS} ingredients and recalls the physical definition of the Rosseland mean opacity. Section~\ref{sec_update} introduces the major updates and revisions implemented in the new version \texttt{\AE SOPUS\,2.0},  and discusses the optimization of the opacity sampling method. Section~\ref{sec_results} presents and examines the results, with particular focus on Rosseland mean opacities computed for scaled-solar abundances.  Section~\ref{sec_evtest} explores first tests of the new opacities in stellar models.
Section~\ref{sec_opactables} introduces the new opacity tables, accessible via a public repository, while Section~\ref{sec_web}  advertises the web-interface for on-the-fly opacity computation.
Finally, Section~\ref{sec_conclu} concludes the paper.

\section{Equation of state and opacity in {\AE SOPUS}}
\label{sec_aesopus}
For a detailed description of the \texttt{\AE SOPUS} code, see \citet[][initial version \texttt{\AE SOPUS\,1.0}]{Marigo_Aringer_2009}. Suffice it here to recall the basic ingredients.
\texttt{\AE SOPUS} solves the equation of state for more than 800 species (about 300 atoms and ions, and 500 molecules) in the gas phase, under the assumption of an ideal gas in both thermodynamic and instantaneous chemical equilibrium.
For all elements from C to Ni, we take into account ionization stages from I to V, (up to VI for O and Ne), and for heavier atoms from Cu to U, we consider ionization stages from I to III.
\texttt{\AE SOPUS} accounts for continuum opacity processes
(Rayleigh scattering, Thomson scattering,
bound-free absorption due to photoionization,
free-free absorption, collision-induced absorption),
and line opacity processes (atomic bound-bound absorption, and molecular band absorption).
For a description of the opacity sources in the current version of \texttt{\AE SOPUS}, referred to as \texttt{\AE SOPUS\,2.0}, see 
Tables \ref{tab_opacsource} and \ref{tab_opacmol} below.

\subsection{The Rosseland Mean Opacity}
\label{ssec_ross}
The solution to the radiation transfer equation greatly simplifies in a gas under conditions of local thermodynamic equilibrium, energy transport diffusion approximation, and spherical symmetry, such that the flux $F(r)$ at the radial coordinate $r$, with gas density $\rho$ and temperature $T$, becomes 
\begin{equation}
F(r)=-\displaystyle\frac{4\pi}{3}\frac{1}{\kappa_{\rm
R}(\rho,T)}\displaystyle\frac{\partial B(r,\,T)}{\partial r}
\end{equation}
where $B(r,\,T)$ is the integral of the
Planck function over frequency, and \kR\ is a frequency-integrated average opacity known as the Rosseland mean opacity, which is calculated as follows.

For any given combination $(\rho, T)$,  we first compute the total monochromatic opacity cross section per unit mass (in cm$^2$ g$^{-1}$), by adding all the contributions from true absorption and scattering
\begin{eqnarray}
\kappa(\nu) & = & \displaystyle\sum_{j}[\kappa_j^{\rm abs}(\nu) + \kappa_j^{\rm scatt}(\nu)] \\
& = & \displaystyle\sum_{j} \frac{n_j}{\rho}\,[\sigma_j^{\rm abs}(\nu)\, 
(1-e^{-h\nu/k_{\rm B}T}) + \sigma_j^{\rm scatt}(\nu)]\,,
\end{eqnarray}
where  $n_j$ is the number density of particles of type $j$, $\sigma_j^{\rm abs/scatt}(\nu)$ is the corresponding absorption/scattering monochromatic cross section (in cm$^2$),  and the factor
$(1-e^{-h\nu/k_{\rm B}T})$ accounts for stimulated emission.

Second, we integrate over frequency to obtain the Rosseland mean opacity, $\kappa_{\rm R}$ (in cm$^2$ g$^{-1}$):
\begin{equation}
  \frac{1}{\kappa_{\rm R}(\rho,T)} = \displaystyle\frac{\displaystyle\int_0^\infty \displaystyle\frac{1}{\kappa(\nu)}
\frac{\partial B_\nu}{\partial T} d\nu}{\displaystyle\int_0^\infty \displaystyle\frac{\partial B_\nu}{\partial T} d\nu}\,,
\label{eq_rosseland}
\end{equation}
which is a harmonic weighted average,  with weights equal to the temperature derivatives of the Planck distribution with respect to temperature, $\displaystyle\frac{\partial B_\nu}{\partial T}$.

For ease of use, Rosseland mean opacity tables are typically constructed as a function of the logarithm of the temperature $T$ (in K units),  and the logarithm of the $R$ variable, which is defined
as $R= \rho\, T_6^{-3}$ (with $\rho$ in g\,cm$^{-3}$ and $T_6=T/(10^6\,{\rm K})$).
Employing the $R$ parameter rather than density $\rho$ or pressure $P$ allows the opacity tables to cover rectangular regions of the $(R,T)$-plane and provides a suitable format for smooth opacity interpolation.
Our Rosseland mean opacity tables extend over the temperature range $3.2 \le \log(T) \le 4.5 $, and the $R$ interval $-1.0 \le \log(R) \le 8.0$.

\begin{table}
\begin{center}
\caption{Scattering and Absorption Processes\label{tab_opacsource}}
\begin{tabular}{l|l|l|l}
\hline
\hline
Process  & Symbol & Reaction & References and Comments \\
\hline
\multirow{3}{*}{\em Rayleigh} & $\sigma_{\rm Ray}$($\HH_2$) & $\HH_2 + h\nu
\rightarrow \HH_2 + h\nu'$ & \cite{DalgarnoWilliams_62} \\
& $\sigma_{\rm Ray}$($\HH$) & $\HH + h\nu \rightarrow \HH + h\nu'$&
\multirow{1}{*}{\cite{Gavrila_67} using fit of \cite{Ferland_00}}\\
& $\sigma_{\rm Ray}$($\He$) & $\He + h\nu \rightarrow \He + h\nu'$& 
\multirow{1}{*}{\cite{Dalgarno_62}}\\
\hline
{\em Thomson} & Th($\e$) &$\e + h\nu \rightarrow \e + h\nu'$ & NIST
(2018 CODATA recommended value)\\
\hline
\multirow{17}{*}{\em free-free} & $\sigma_{\rm ff}$($\HH^-$) & 
$\HH+\e +h\nu \rightarrow \HH + \e$& \cite{Bell_Berrington_87} using fit of \cite{John_88} \\
& \multirow{1}{*}{$\sigma_{\rm ff}$($\HH$)} & 
\multirow{1}{*}{$\HH^+ + \e + h\nu \rightarrow \HH^+ + \e $} & Method as \cite{Kurucz_70} based on \cite{Karzas_Latter_61}\\
&$\sigma_{\rm ff}$($\HH_2^+$) & $\HH^+ + \HH + h\nu \rightarrow \HH^+ + \HH$ & \cite{Lebedev_etal_03}\\
&$\sigma_{\rm ff}$($\HH_2^-$) & $ \HH_2 + \e +h\nu \rightarrow \HH_2 + \e$ & \cite{John_75}\\
&$\sigma_{\rm ff}$($\HH_3$) & $\HH_3^+ + \e + h\nu \rightarrow \HH_3^+ + \e $& $\sigma_{\rm ff}(\HH_3)=\sigma_{\rm ff}(\HH)$ (assumed) \\
&$\sigma_{\rm ff}$($\He^-$) & $\He + \e + h\nu \rightarrow \He+ \e$ & \cite{John_94} \\
&$\sigma_{\rm ff}$($\He$) &$\He^+ + \e + h\nu \rightarrow \He^+ + \e$& $\sigma_{\rm ff}(\He)=\sigma_{\rm ff}(\HH)$ (assumed)\\
&$\sigma_{\rm ff}$($\He^+$) &$\He^{++} + \e + h\nu \rightarrow \He^{++} + \e$ & $\sigma_{\rm ff}(\He^+)=\sigma_{\rm ff}(\HH)$ (assumed)\\
&$\sigma_{\rm ff}$(${\rm Li}^-$) &${\rm Li} + \e + h\nu \rightarrow {\rm Li}+ \e$ &    \cite{RamsbottomBell_96}\\
&$\sigma_{\rm ff}$(${\rm C}^-$) &${\rm C} + \e + h\nu \rightarrow {\rm C}+ \e$ & \cite{Bell_etal_88}\\
&$\sigma_{\rm ff}$(${\rm N}^-$) &${\rm N} + \e + h\nu \rightarrow {\rm N}+ \e$ & \cite{Ramsbottom_etal_92}\\
&$\sigma_{\rm ff}$(${\rm O}^-$) &${\rm O} + \e + h\nu \rightarrow {\rm O}+ \e$ & \cite{John_75}\\
&$\sigma_{\rm ff}$(${\rm Ne}^-$) &${\rm Ne} + \e + h\nu \rightarrow {\rm Ne}+ \e$ & \cite{John_96}\\
&$\sigma_{\rm ff}$(${\rm Cl}^-$) &${\rm Cl} + \e + h\nu \rightarrow {\rm Cl}+ \e$ & \cite{JohnMorgan_75}\\
&$\sigma_{\rm ff}$(${\rm H_2O}^-$) &${\rm H_2O} + \e + h\nu \rightarrow {\rm H_2O}+ \e$ & \cite{John_75}\\
&$\sigma_{\rm ff}$(${\rm CO}^-$) &${\rm CO} + \e + h\nu \rightarrow {\rm CO}+ \e$ & \cite{John_75}\\
&$\sigma_{\rm ff}$(${\rm N_2}^-$) &${\rm N_2} + \e + h\nu \rightarrow {\rm N_2}+ \e$ & \cite{John_75}\\
\hline
\multirow{7}{*}{\em bound-free} & $\sigma_{\rm bf}$($\HH^-$) & $\HH^- + h\nu \rightarrow \HH + \e$ & \cite{McLaughlin_2017}\\ 
&  \multirow{2}{*}{$\sigma_{\rm bf}$($\HH$)} & \multirow{2}{*}{$\HH + h\nu \rightarrow \HH^+ + \e$} & Method as in \cite{Kurucz_70} based on\\
& & & \cite{Gingerich_64} and \cite{Karzas_Latter_61}\\\cline{4-4}
 & $\sigma_{\rm bf}$($\HH_2^+$) & $\HH_2^+ + h\nu \rightarrow \HH^+ + \HH $ & \cite{Lebedev_etal_03}\\\cline{4-4}
&\multirow{2}{*}{$\sigma_{\rm bf}$($\He$)} &  \multirow{2}{*}{$\He + h\nu \rightarrow \He^+ + \e$} & Method as in \cite{Kurucz_70} based on\\
& & & \cite{Gingerich_64} and \cite{HungerBlerkom_67} \\\cline{4-4}
& $\sigma_{\rm bf}$($\He^+$) & $\He^+ + h\nu \rightarrow \He^{++} + \e$ &  \cite{HungerBlerkom_67}\\
\hline
\multirow{1}{*}{\em bound-bound} & $\sigma_{\rm bb}$($\HH$) & \multirow{1}{*}{$\HH + h\nu \rightarrow \HH^*$}& \cite{Kurucz_70} including Stark broadening\\
\hline
\multirow{8}{2cm}{\em Collision\\induced\\absorption} & \multirow{2}{*}{$\sigma_{\rm CIA}$($\HH_2/\HH_2$)} & \multirow{2}{*}{$\HH_2+\HH_2+h\nu \rightarrow \HH_2+\HH_2$} & $200\K<T<3000 \K$,  $20\cm^{-1}<\tilde{\nu}<10 000\cm^{-1}$ \\
& & & \cite{Abel_etal_11} \\\cline{4-4}
 & \multirow{2}{*}{$\sigma_{\rm CIA}$($\HH_2/\HH$)} & \multirow{2}{*}{$\HH_2+\HH+h\nu \rightarrow \HH_2+\HH$} & $1000\K<T<2500\K$, $100\cm^{-1}<\tilde{\nu}<10000\cm^{-1}$ \\
& & & \cite{GustafssonFrommhold_03}\\\cline{4-4}
 & \multirow{2}{*}{$\sigma_{\rm CIA}$($\HH_2/\He$)} & \multirow{2}{*}{$\HH_2+\He+h\nu \rightarrow \HH_2+\He$} & $200\K<T<9900\K$, $20\cm^{-1}<\tilde{\nu}<20000\cm^{-1}$ \\
& & & \cite{Abel_etal_12}\\\cline{4-4}
& \multirow{2}{*}{$\sigma_{\rm CIA}$($\HH/\He$)} & \multirow{2}{*}{$\HH+\He+h\nu \rightarrow \HH+\He$} & $1500\K <T<10000 \K$, $50\cm^{-1}<\tilde{\nu}<11000\cm^{-1}$\\
& & &  \cite{Gustafsson_etal_01} \\
\hline
\multirow{5}{2cm}{\em bound-free\\free-free} & {C, N, O} &  \multirow{5}{3cm}{${\rm X} + h\nu \rightarrow {\rm X}^+ + \e$\\${\rm X}+\e +h\nu \rightarrow {\rm X} + \e$} \\
& {Ne, Na, Mg} & \\
& {Al, Si, S} &  & Opacity Project: \cite{OP_95}   for $\log(T) \ge 3.6$\\
& {Ar, Ca, Cr} &  \\
& {Mn, Fe, Ni} & \\
\hline
\multirow{3}*{\em bound-free} & \multirow{1}*{CI, NI} & \multirow{3}{*}{${\rm X} + h\nu \rightarrow {\rm X}^+ + \e$}& Method as in \cite{Kurucz_70} based on  \cite{Peach_70} \\
& {OI, MgI} & &and \cite{Henry_70} for $\log(T) < 3.6$\\
&{AlI, SiI} & & \\
\hline
\end{tabular}
\end{center}
\tablecomments{X denotes the generic atom/ion. Molecular absorption sources are described in Table~\ref{tab_opacmol}.}
\end{table}
\begin{table}
\begin{center}
\caption{Spectral Line Data for Molecular Absorption \label{tab_opacmol}}
\begin{tabular}{clcl}
\hline
\hline
\noalign{\smallskip}
{Molecule} & \multicolumn{1}{l}{Reference} & Molecule &  \multicolumn{1}{l}{Reference}\\
\hline
 HF &  \cite{HF_COXON2015133,HF_LI201378} & CaO & \cite{CaO_10.1093/mnras/stv2858} \\
 HCl &  \cite{HCL_LI20111543}& CH$_3$ & \cite{CH3_2019JPCA..123.4755A} \\
 CH &  \cite{CH_Masseron_etal_2014} &CH$_3$Cl & \cite{CH3CL_10.1093/mnras/sty1542}\\ 
 C$_2$ &  \cite{C2_10.1093/mnras/staa1954,C2_10.1093/mnras/sty2050} & CP & \cite{CP_QIN2021107352, CP_RAM2014107} \\
 CN &   \cite{CN_10.1093/mnras/stab1551}& H$_2$ & \cite{H2_2019AA...630A..58R} \\
 CO &   \cite{CO_doi:10.1063/5.0063256, CO_Li_2015} & H$_2$S & \cite{H2S_10.1093/mnras/stw1133}\\
 OH &   \cite{OH_YOUSEFI2018416,OH_BROOKE2016142} & KCl & \cite{NaCl_KCl_10.1093/mnras/stu944}\\
 SiO &   \cite{SiO_2022MNRAS.510..903Y}& KF & \cite{NaF_KF_FROHMAN2016104} \\
 TiO &   \cite{TiO_10.1093/mnras/stz1818} & KOH & \cite{KOH_NaOH_10.1093/mnras/staa4041}\\
 VO &   \cite{VO_10.1093/mnras/stw1969} & LiCl & \cite{LiCl_Bittner_2018}\\
 CrH &  \href{http://bernath.uwaterloo.ca}{Diatomic Database of P.F.~Bernath}& MgF & \cite{MgF_HOU2017511} \\
 FeH &  \cite{FeH_Dulick_2003} & MgO & \cite{MgO_10.1093/mnras/stz912}\\
 YO &  \cite{YO_C9CP03208H} & N$_2$& \texttt{HITRAN}:  \cite{HITRAN2020_GORDON2022107949} \\
 ZrO &  \cite{VanEck_etal_17, Plez_2012} &NaCl & \cite{NaCl_KCl_10.1093/mnras/stu944} \\ H$_2$O &  \cite{H2O_10.1093/mnras/sty1877} & NaF & \cite{NaF_KF_FROHMAN2016104}\\
 HCN &  \cite{HCN_10.1111/j.1365-2966.2005.09960.x} & NaO & \cite{NaO_2022MNRAS.511.2349M} \\
 C$_3$ &   \cite{Jorgensen_etal_89} & NaOH & \cite{KOH_NaOH_10.1093/mnras/staa4041} \\
 CO$_2$ &  \cite{CO2_10.1093/mnras/staa1874} & NH$_3$ & \cite{NH3_10.1093/mnras/stz2778, NH3_ALDERZI2015117} \\
 SO$_2$ &   \cite{SO2_10.1093/mnras/stw849}& NO & \cite{NO_10.1093/mnras/stab1154}\\
 C$_2$H$_2$ &   \cite{C2H2_10.1093/mnras/staa229} & NS & \cite{NS_10.1093/mnras/sty939}\\
 AlH &  \cite{AlH_2018MNRAS.479.1401Y}& PH & \cite{PH_10.1093/mnras/stz1856} \\
 AlO &  \cite{AlO_2021MNRAS.508.3181B,AlO_10.1093/mnras/stv507} & PH$_3$ & \cite{PH3_10.1093/mnras/stu2246}\\
 CaH &  \cite{MgH_CaH_2022MNRAS.511.5448O}
 & PN & \cite{PN_10.1093/mnras/stu1854}\\
 CH$_4$ & \cite{CH4_2017AA...605A..95Y, CH4_10.1093/mnras/stu326} & PO & \cite{PO_PS_10.1093/mnras/stx2229}\\
 CS &  \cite{CS_10.1093/mnras/stv1543} & PS & \cite{PO_PS_10.1093/mnras/stx2229}\\
 LiH &  \cite{LiH_10.1111/j.1365-2966.2011.18723.x}& ScH & \cite{ScH_2021AA...646A..21C, ScH_2015MolPh.113.1998L}\\
 MgH &  \cite{MgH_CaH_2022MNRAS.511.5448O}& SiH$_4$ & \cite{SiH4_10.1093/mnras/stx1952} \\
 TiH &  \cite{TiH_Burrows_2005} & SiO$_2$ & \cite{SiO2_10.1093/mnras/staa1287}\\
 NaH &  \cite{NaH_2015MNRAS.451..634R} & SiS & \cite{SiS_10.1093/mnras/sty998}\\
 NH &  \cite{NH_FERNANDO201829, NH_doi:10.1063/1.4923422}& LiF & \cite{LiF_Bittner_2018} \\
 SH &  \cite{SH_10.1093/mnras/stz2517}& O$_2$& \cite{O2_2021AA...646A..21C, HITRAN2016_GORDON20173} \\
 SiH &  \cite{SiH_10.1093/mnras/stx2738} & OCS & \cite{HITRAN2020_GORDON2022107949} \\
 AlCl &  \cite{MOLLIST_BERNATH2020106687} & H$_3^+$ & \cite{H3p_10.1093/mnras/stx502} \\
 AlF &  \cite{MOLLIST_BERNATH2020106687} &H$_3$O$^+$ & \cite{H3Op_10.1093/mnras/staa2034} \\
 BeH &  \cite{BeH_Darby_Lewis_2018}& HeH$^+$& \cite{HeHp_Amaral_2019}\\
 C$_2$H$_4$ &  \cite{C2H4_10.1093/mnras/sty1239} &LiH$^+$ & \cite{LiHp_10.1111/j.1365-2966.2011.18723.x}\\
 CaF &  \cite{CaF_HOU201844} &OH$^+$ & \cite{OHp_Hodges_2017}\\
 CS$_2$ & \texttt{HITRAN}:   \cite{HITRAN2020_GORDON2022107949} & SO& \texttt{HITRAN}: \cite{HITRAN2020_GORDON2022107949}\\
 HI  & \texttt{HITRAN}:  \cite{HITRAN2020_GORDON2022107949} & ClO & \texttt{HITRAN}: \cite{HITRAN2020_GORDON2022107949}\\
 HBr & \texttt{HITRAN}: \cite{HITRAN2020_GORDON2022107949} & O$_3$ & \texttt{HITRAN}: \cite{HITRAN2020_GORDON2022107949}\\
\hline
\end{tabular}
\end{center}
\tablecomments{
Most of the monochromatic absorption cross sections (but for C$_3$ and ZrO) are calculated using the \texttt{EXOCROSS} tool, available in the \href{https://www.ExoMol.com/data/molecules/}{ExoMol} database, from the corresponding line lists. Line broadening accounts for thermal Doppler and micro-turbulent velocity.}
\end{table}

\section{Major updates in \AE SOPUS\,2.0} \label{sec_update}
In this work we expand and update a significant number of opacity sources (Tables~\ref{tab_opacsource} and \ref{tab_opacmol}) and thermodynamic data. 
In addition, we revise various partition functions for diatomic molecules, taken from \cite{Barklem_Collet_2016}, and from \texttt{ExoMol} database \citep{EXOMOL_2012MNRAS.425...21T}.
Here below, we only highlight the most significant changes in opacity, that refer to the continuous absorption from the negative hydrogen ion, collision-induced absorption, and molecular absorption. In addition to H$^-$, negative ion free-free opacity from other species is also included and/or revised (He$^-$, Li$^-$, C$^-$, N$^-$, O$^-$, Ne$^-$, Cl$^-$, H$_2$O$^-$, CO$^-$, N$_2^-$).

\subsection{Photodetachment of H$^-$}
Since the pioneering work of \cite{ChandrasekharBreen_46}, continuous absorption from the negative hydrogen ion has been recognized as an important opacity source in the stellar atmospheres. \cite{John_88} analytic fits to theoretical data for the free-free \citep{Bell_Berrington_87} and bound-free \citep[][for $\lambda < 1.6419\, \mu $m]{Wishart_79} cross sections are a classic reference study of H$^-$, that is used in most opacity codes.

\begin{figure}[h!]
    \centering
    \includegraphics[width=0.48\textwidth]{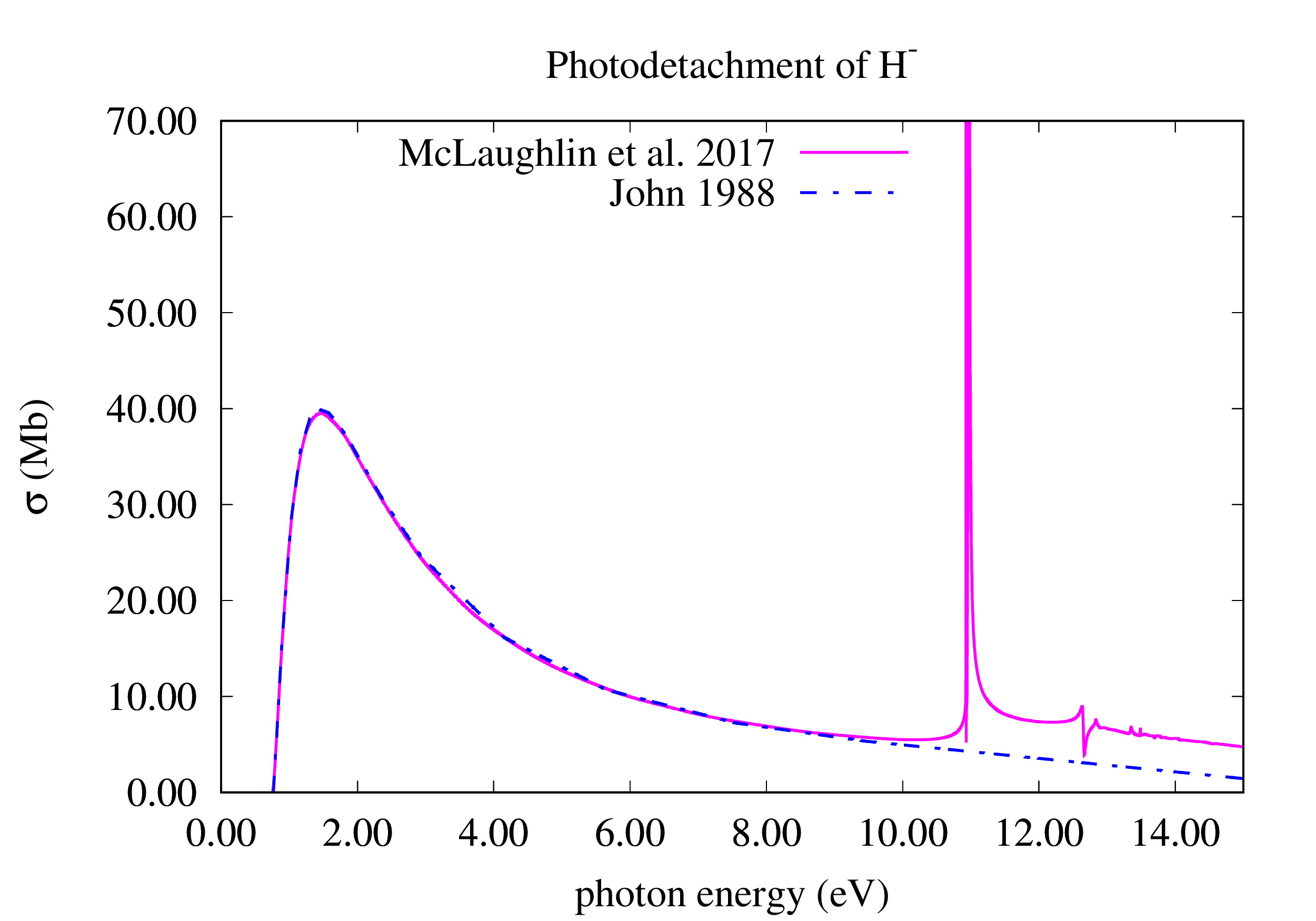}
    \caption{Photodetachment cross section from \citet[][solid line]{McLaughlin_2017},  compared to \citet{John_88}'s fit of the cross section data from \citet[][dashed line]{Wishart_79}. As can be seen, the latter does not account for auto-detaching resonances at photon energies above 10 eV. The cross section $\sigma$ is in units of $10^6$ barn (Mb). }
    \label{fig_Hmin}
\end{figure}

In this work we base on the recent study carried out by \cite{McLaughlin_2017} to revise the photodetachment cross section of H$^-$.
\cite{McLaughlin_2017} combine R-matrix calculations and comparison to available experimental data to build an H$^-$ photodetachment cross section that is accurate over a wide range of photon energies and takes into account a series of auto-detaching shape and Feshbach resonances at photon energies ranging from  $10.92$ to $14.35$ eV. As discussed by \cite{McLaughlin_2017} and shown in Figure~\ref{fig_Hmin}, the simple fit to \cite{Wishart_79} calculations cannot reproduce the behavior of the cross section in the region of the auto-detaching resonances beyond 10 eV.

\subsection{Collision-Induced Absorption}
Collision-induced absorption (CIA) is caused by collisions of molecules and atoms in a gas of relatively high density \citep{Frommhold_1994}.
\begin{figure}
    \centering
    \includegraphics[width=0.48\textwidth]{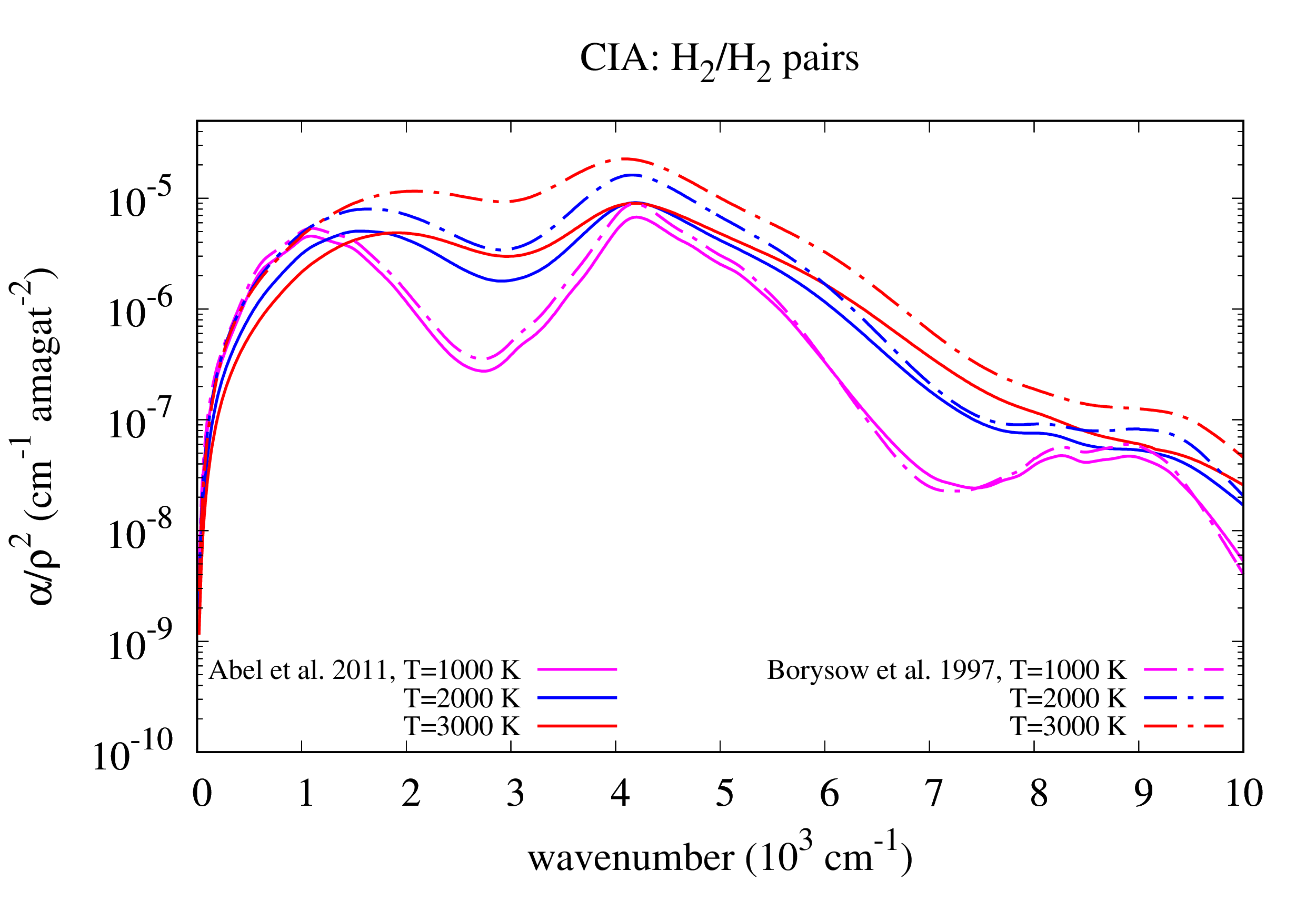}
    \includegraphics[width=0.48\textwidth]{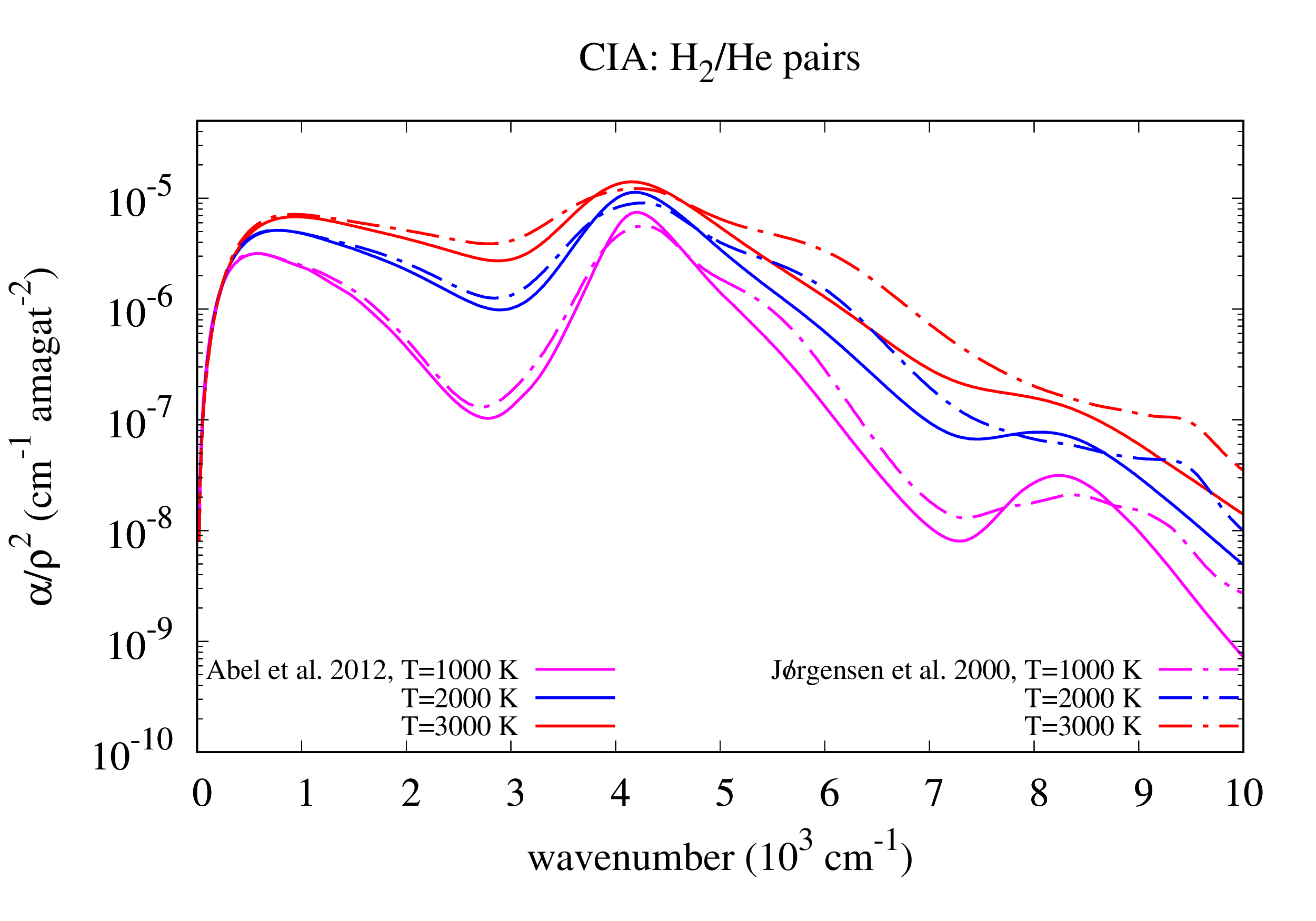}
    \caption{Collision-induced absorption of H$_2$/H$_2$ and H$_2$/He pairs. Following standard notation, the absorption coefficient $\alpha$ is normalized by gas density squared.
    The data adopted in the current version
     \texttt{\AE SOPUS\,2.0} (solid line; \cite{Abel_etal_11} for H$_2$/H$_2$, and \cite{Abel_etal_12} for H$_2$/He) is compared to that used in \texttt{\AE SOPUS\,1.0} (dash-dot line; \cite{Borysow_etal_97} for H$_2$/H$_2$, and \cite{Jorgensen_etal_00} for H$_2$/He), for three temperature values.}
    \label{fig_ciah2h2}
\end{figure}
Because hydrogen and helium gases dominate the atmospheres of giant stars and brown dwarfs, collision complexes such as H$_2$/H$_2$, H$_2$/H, H$_2$/He, H/He  may  contribute significantly to absorption in these layers. 
In this work we adopt the latest parametrization set up in \texttt{HITRAN} spectroscopic database \citep{Karman_etal_2019}. In particular, we adopt the results from \cite{Abel_etal_11} for 
H$_2$/H$_2$, \cite{GustafssonFrommhold_03} for 
H$_2$/H, \cite{Abel_etal_12} for H$_2
$/He, and \cite{Gustafsson_etal_01} for H/He.

As discussed by \cite{Abel_etal_11} and shown in Figure~\ref{fig_ciah2h2} (left panel), the H$_2$ rotational fundamental band and first-overtone structures at the lowest temperature ($T=1000$ K) are particularly pronounced, but as temperature rises, the inter-band minima shrink and the H$_2$ bands blend more and more.
There are substantial differences between the results of \cite{Abel_etal_11} and the earlier calculations of \cite{Borysow_etal_97}, especially for higher temperatures ($T> 1000$ K). The inconsistencies are most likely due to a less accurate characterization of the induced dipole surface in older CIA studies, the results of which are adopted in opacity calculations by \cite{Marigo_Aringer_2009} and \cite{Ferguson_etal_05}. Similar considerations apply to  H$_2$/He collision induced absorption (Figure~\ref{fig_ciah2h2}, right panel). 

\subsection{Molecular Absorption}
\label{ssec_molabs}
This work represents a substantial advancement over \cite{Marigo_Aringer_2009} in terms of molecular absorption (see Table~\ref{tab_opacmol}). We extend the number of absorbing molecules to 80 (in \citealt[][\texttt{\AE SOPUS\,1.0},]{Marigo_Aringer_2009} there were 20), and we carry out a systematic update of the monochromatic cross sections, $\sigma_j(\nu)$. 
The update is primarily based on the \texttt{ExoMol} line list database \citep{EXOMOL_2012MNRAS.425...21T} and its public tools, in particular the software \texttt{EXOCROSS} to compute the absorption cross sections \citep{EXOCROSS_2018}. Data from \texttt{HITRAN} are also included \citep{HITRAN2020_GORDON2022107949}.

Table~\ref{tab_opacmol} contains the complete record of absorbing molecules, together with the corresponding line list sources. For each molecular species included in our code the monochromatic cross section, $\sigma_j(\nu)$, is taken from opacity sampling (OS) files produced for a selected frequency grid, that are calculated directly from the corresponding line list.

\begin{figure*}
    \centering
    \includegraphics[width=0.48\textwidth]{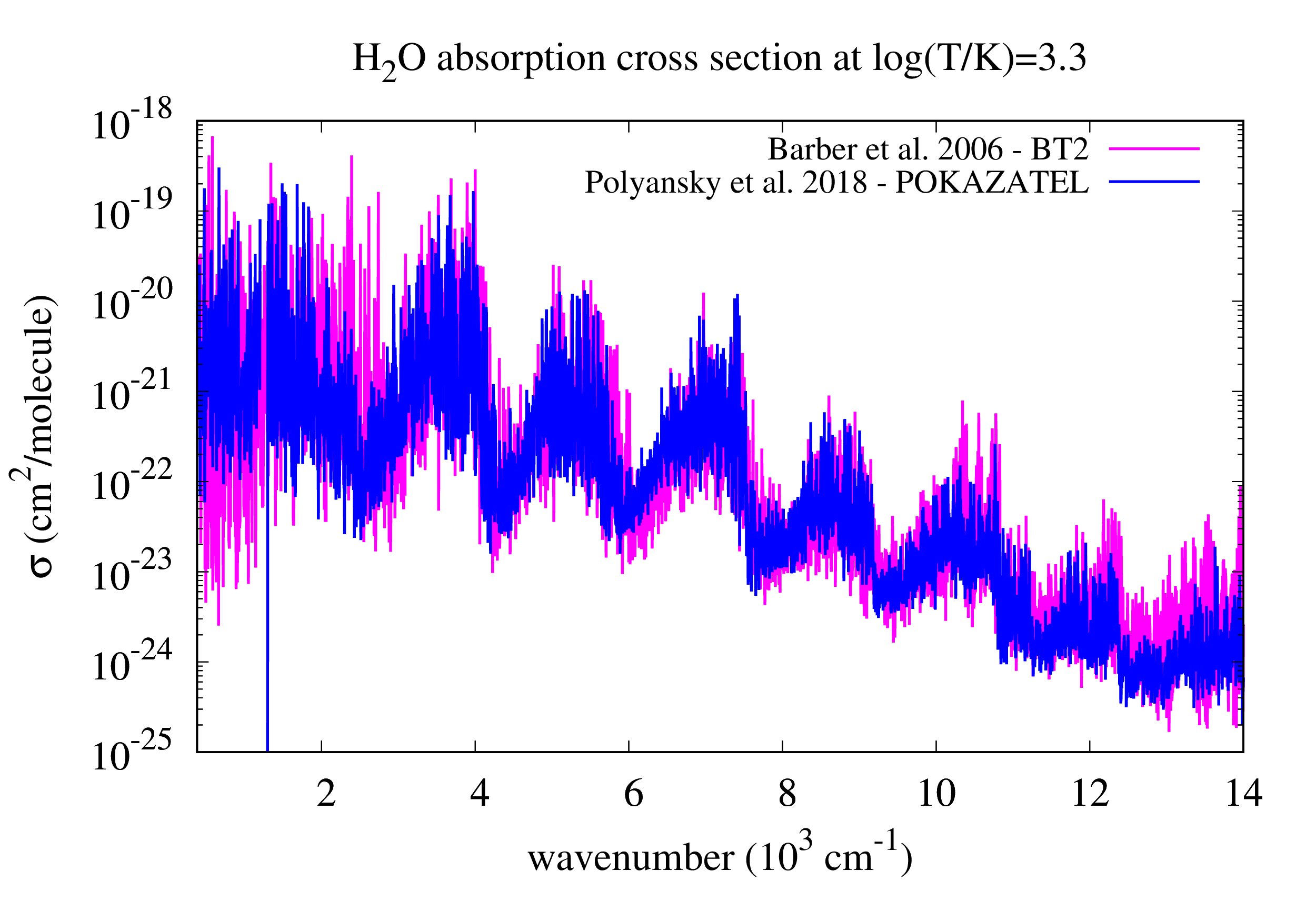}
    \includegraphics[width=0.48\textwidth]{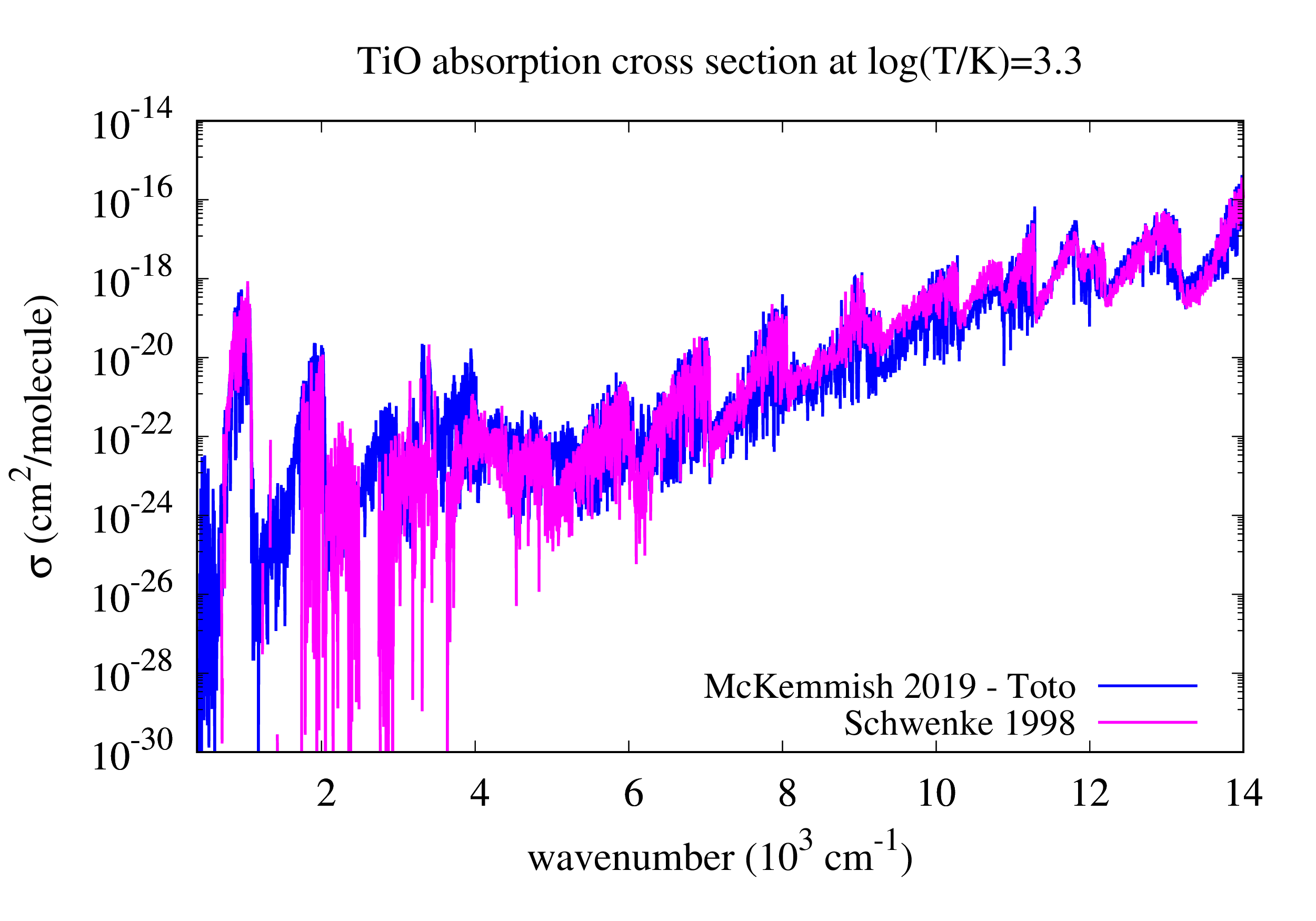}
    \caption{Absorption cross section of two important molecules in the cool atmospheres of O-rich giants and supergiants, namely: H$_2$O (left panel) and TiO (right panel). We compare the \texttt{ExoMol} recommended line lists (blue) adopted in this work with earlier data (magenta) used in \texttt{\AE SOPUS\,1.0}. Line broadening due to thermal Doppler effect and micro-turbulent velocity was used for both molecules' transitions.}
    \label{fig_h2o_tio}
\end{figure*}
As an example, Figure~\ref{fig_h2o_tio} depicts the cross sections of two relevant molecules in the atmospheres of red giants, which are characterized by a photospheric carbon-to-oxygen ratio\footnote{\co\ is the ratio of C to O abundances (in number) at the star's photosphere.} $\co\,<1$. We compare H$_2$O and TiO absorption from different line lists used in \texttt{\AE SOPUS\,1.0} \citep{Marigo_Aringer_2009} and the current version \texttt{\AE SOPUS\,2.0}.

The \texttt{POKAZATEL} line list for water \citep{H2O_10.1093/mnras/sty1877} has nearly $6 \times 10^9$ lines, while \texttt{BT2} \citep{Barber_etal_06} includes $\approx 500\times 10^6$ transitions.
The \texttt{POKAZATEL} line list, in particular, produces much  weaker absorption in the near-UV region than \texttt{BT2}, which is supported by the findings of a recent study of ultraviolet terrestrial atmospheric absorption \citep{Lampel_etal_17}. Furthermore, unlike \texttt{BT2}, \texttt{POKAZATEL} cross sections become progressively flattened with increasing temperature,  as a result of a more comprehensive treatment, including high J states and vibrational hot bands (see \citealt{H2O_10.1093/mnras/sty1877} for a thorough discussion). 
Overall, these differences in monochromatic cross sections may have a significant impact on the resulting Rosseland mean opacities.

Below we will briefly review two specific aspects about our procedure for treating molecular absorption.

\paragraph{Optimized Opacity Sampling}
When computing Rosseland mean opacities, the frequency grid must be carefully chosen to ensure both fast performance and accuracy. As thoroughly discussed in \citet{Marigo_Aringer_2009}, we use the \citet{Helling_Jorgensen_98} algorithm to optimize the frequency distribution in the opacity sampling technique.

As demonstrated by our earlier tests in \citet{Marigo_Aringer_2009}, computing time increases almost linearly  with opacity-sampling frequency grid size, $n_{\rm OSgrid}$, while gain in precision does not, resulting in Rosseland mean opacities that vary only marginally beyond a certain threshold of a few thousands frequency points.
The results presented here are obtained with an opacity-sampling frequency grid containing  $n_{\rm OSgrid}=5365$ points\footnote{\cite{Helling_Jorgensen_98} investigated the reliability of the radiative transfer solution in hydrostatic \texttt{MARCS} models of cool giant star atmospheres as a function of frequency grid size. Moving from $n_{\rm OSgrid}=22432$ to $n_{\rm OSgrid}=5608$ points, the maximum deviation in surface temperature does not exceed 8 K (see their table 1 and Figure 3).}, yielding a favorable accuracy/computing-time ratio.
\begin{figure}[h!]
    \centering
    \includegraphics[width=0.45\textwidth]{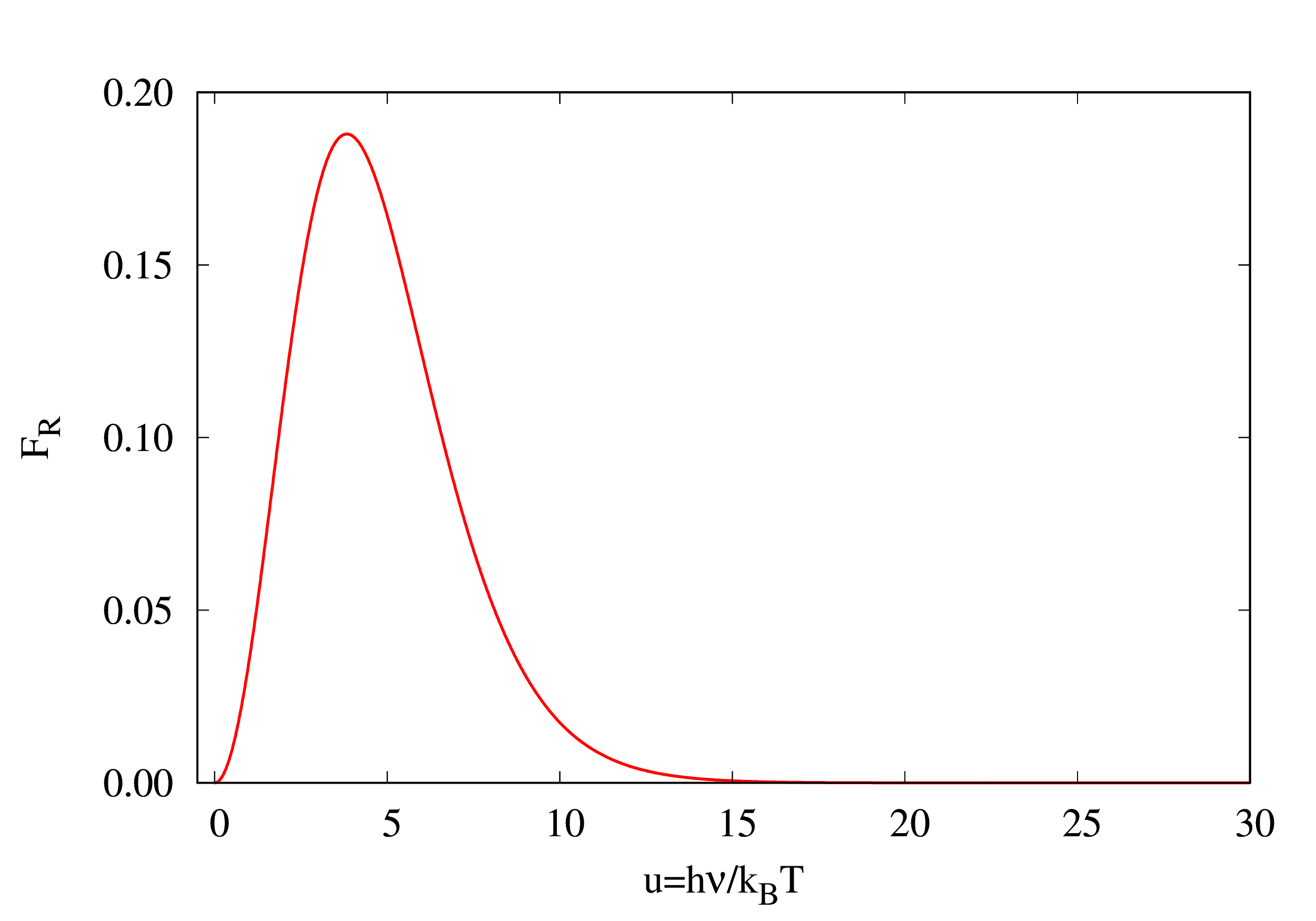}
    \includegraphics[width=0.45\textwidth]{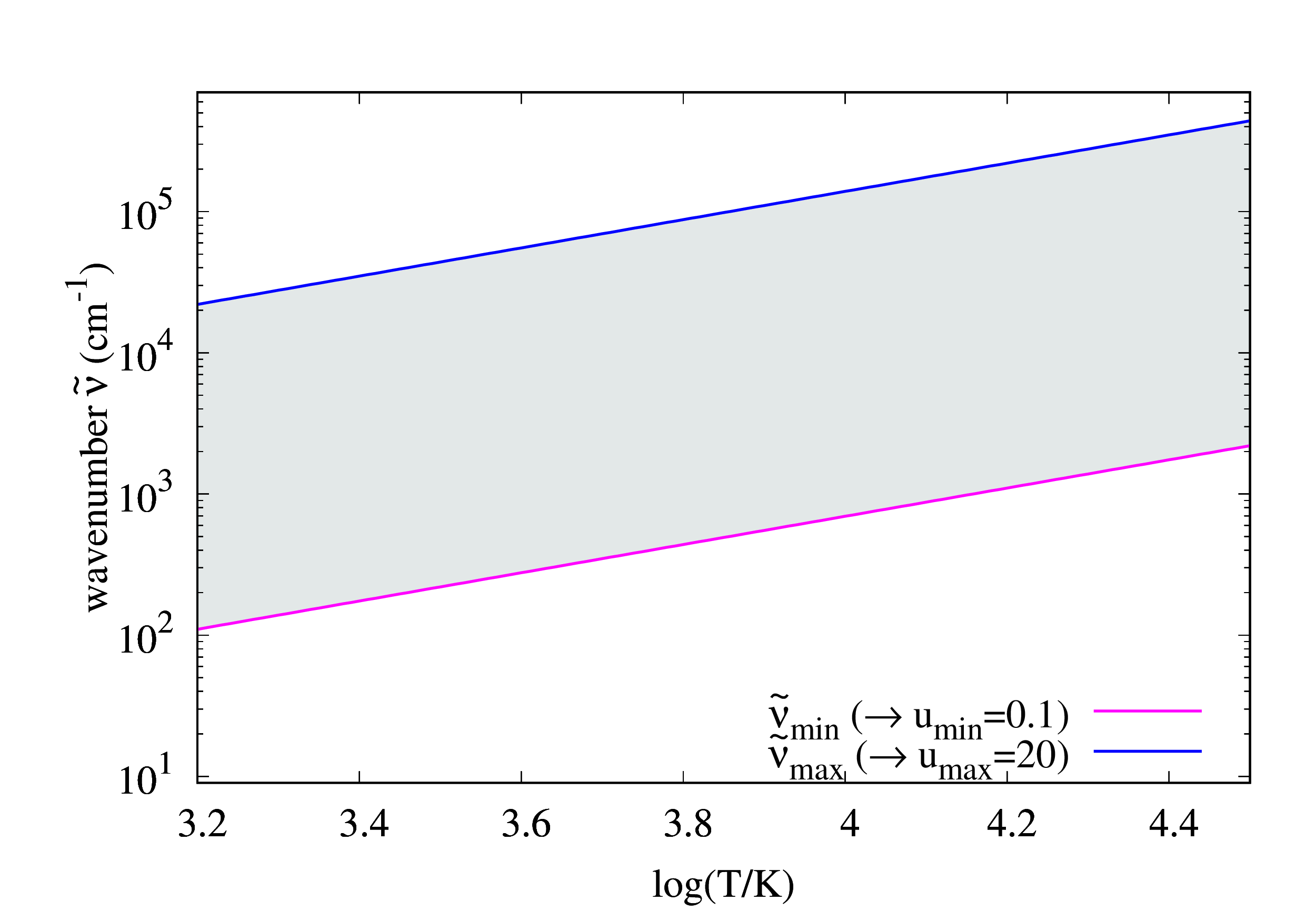}\\
    \caption{{\em Left}: Weighting function, $F_{\rm R}(u)$, in the integral of Equation~(\ref{eq_krossu}). {\em Right}: Dynamic integration limits in wavenumber $\tilde{\nu}$, dependent on temperature, used for the integration of the Rosseland mean opacity. Normalized energies $u_{\rm min}=0.1$ and $u_{\rm max}=20$ correspond to $\tilde{\nu}_{\rm min}$ and $\tilde{\nu}_{\rm max}$.}
    \label{fig_fr}
\end{figure}

In this work we optimize the frequency grid selection even further, by focusing on the lower and upper limits of the integral in  Equation~(\ref{eq_rosseland}), which formally defines the Rosseland mean opacity.
 In \citet{Marigo_Aringer_2009} we used constant integration limits,  corresponding to  wavenumbers $\tilde\nu_{\rm min}=10\, {\rm cm^{-1}}$ and  $\tilde\nu_{\rm max}=2\times10^5\, {\rm cm^{-1}}$. These values adequately cover the frequency range that is relevant for the temperatures under consideration.
We can, however, improve the selection of the integration extremes.
Following \citet{Seaton94} we note that  \kR\ can be easily calculated with:
\begin{equation} 
\label{eq_krossu}
\frac{1}{\kappa_{\rm R}(\rho,T)}=\int_{u_{\rm min}}^{u_{\rm max}}\frac{F_{\rm R}(u)}{\kappa(u)}\rmd  u\, ,
\end{equation}
where the weighting function is expressed as
\begin{equation}
F_{\rm R}(u)=\frac{15}{4\pi^4}\,u^4\exp(-u)/[1-\exp(-u)]^2.
\label{weight}
\end{equation}
Here $\nu$ is the photon frequency, and
$u~=~h\nu/(\kb T)$ is the normalized photon energy. The integration extremes, $u_{\rm min}$ and $u_{\rm max}$, should be chosen in such a way that they vastly encompass the domain where the function $F_{\rm R}(u)$ is not zero.

As shown in Figure~\ref{fig_fr} (left panel), setting $u_{\rm min}=0.1$ and $u_{\rm max}=20$ satisfies this requirement, as these correspond to $\simeq 0.001$ and $\simeq 99.998$ percentiles of $F_{\rm R}(u)$, respectively. Because the normalized energy $u$ varies with temperature, we can make the integration extremes dynamic (see Figure~\ref{fig_fr}, right panel), rather than keeping them fixed for any value of $T$. In this way, we can eliminate unnecessary frequency points that are outside the range $[u_{\rm min},u_{\rm max}]$, which has the added benefit of shortening the numerical integration. This will be especially important for speeding up on-the-fly opacity computations through our public web-interface. 
\begin{figure*}[h!]
    \centering
    \includegraphics[width=0.48\textwidth]{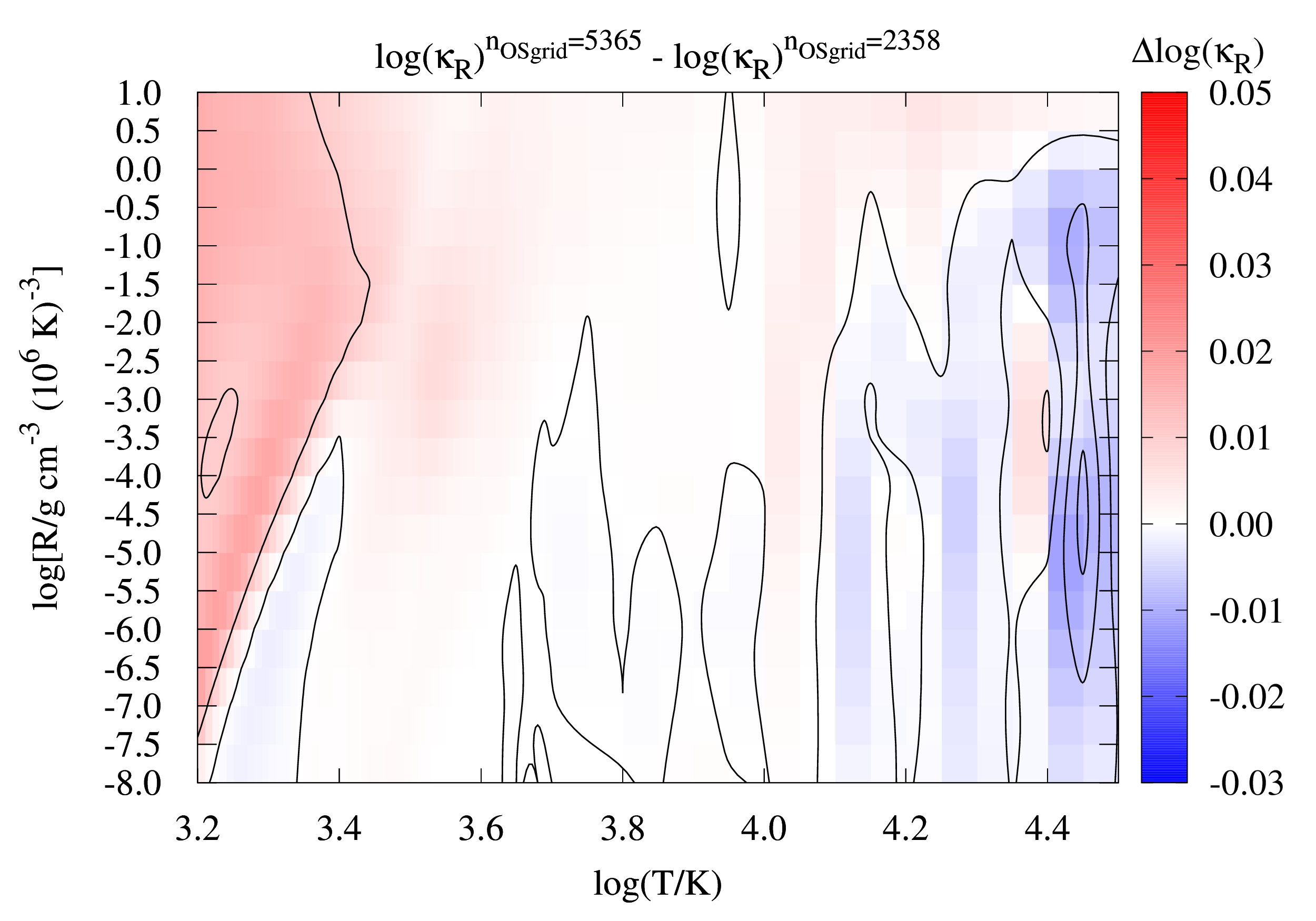}
    \includegraphics[width=0.48\textwidth]{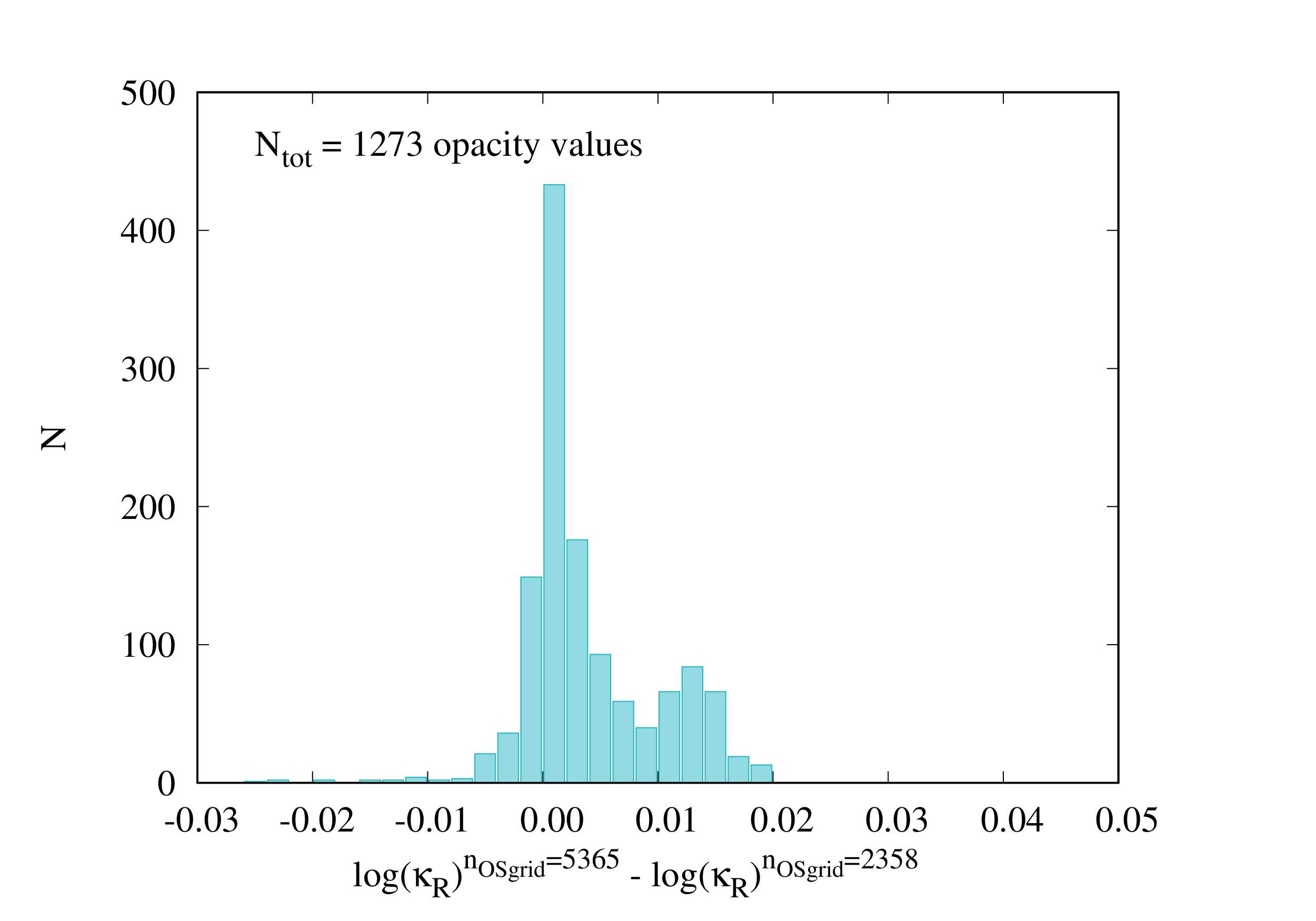} \\
    \includegraphics[width=0.48\textwidth]{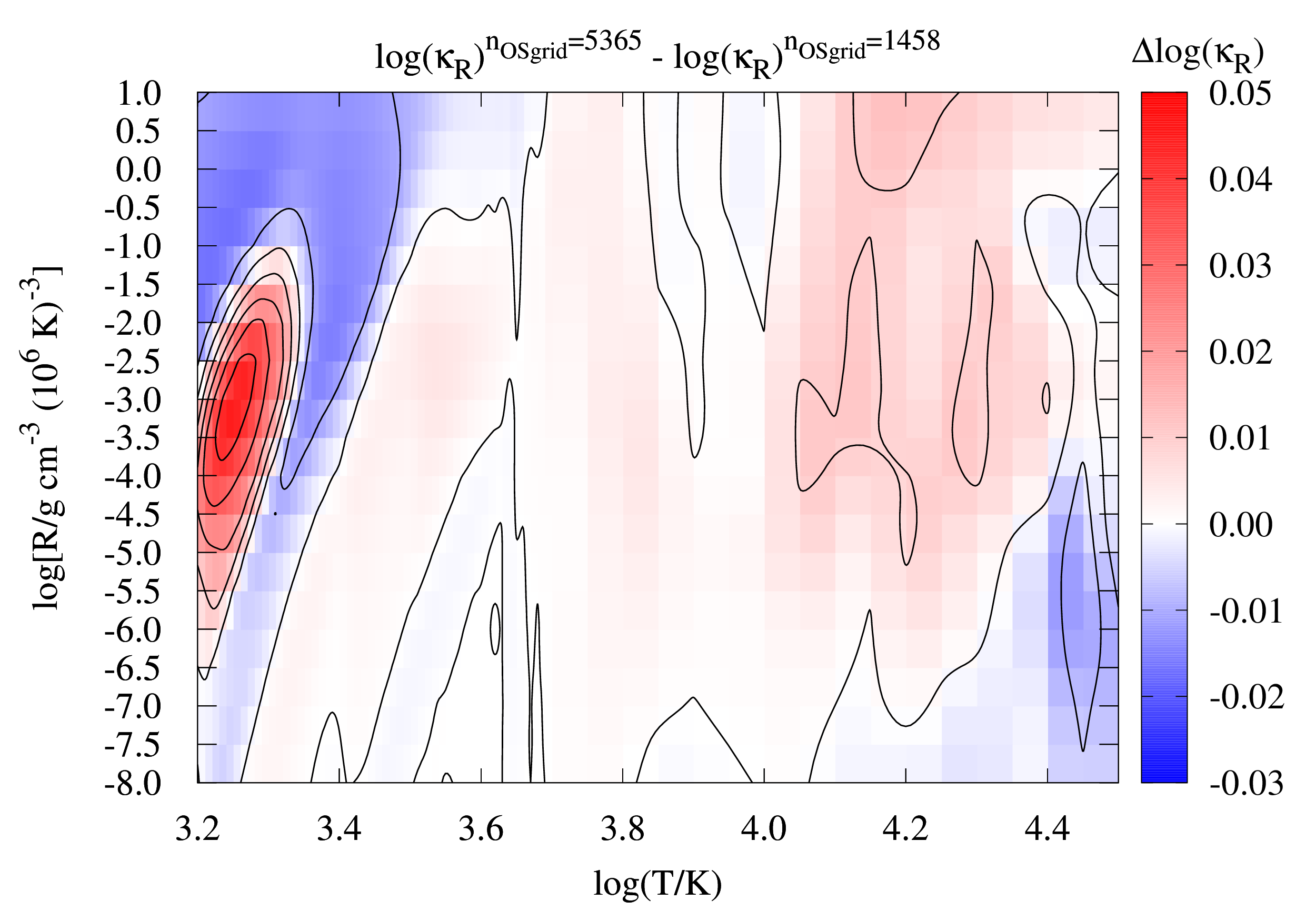}
    \includegraphics[width=0.48\textwidth]{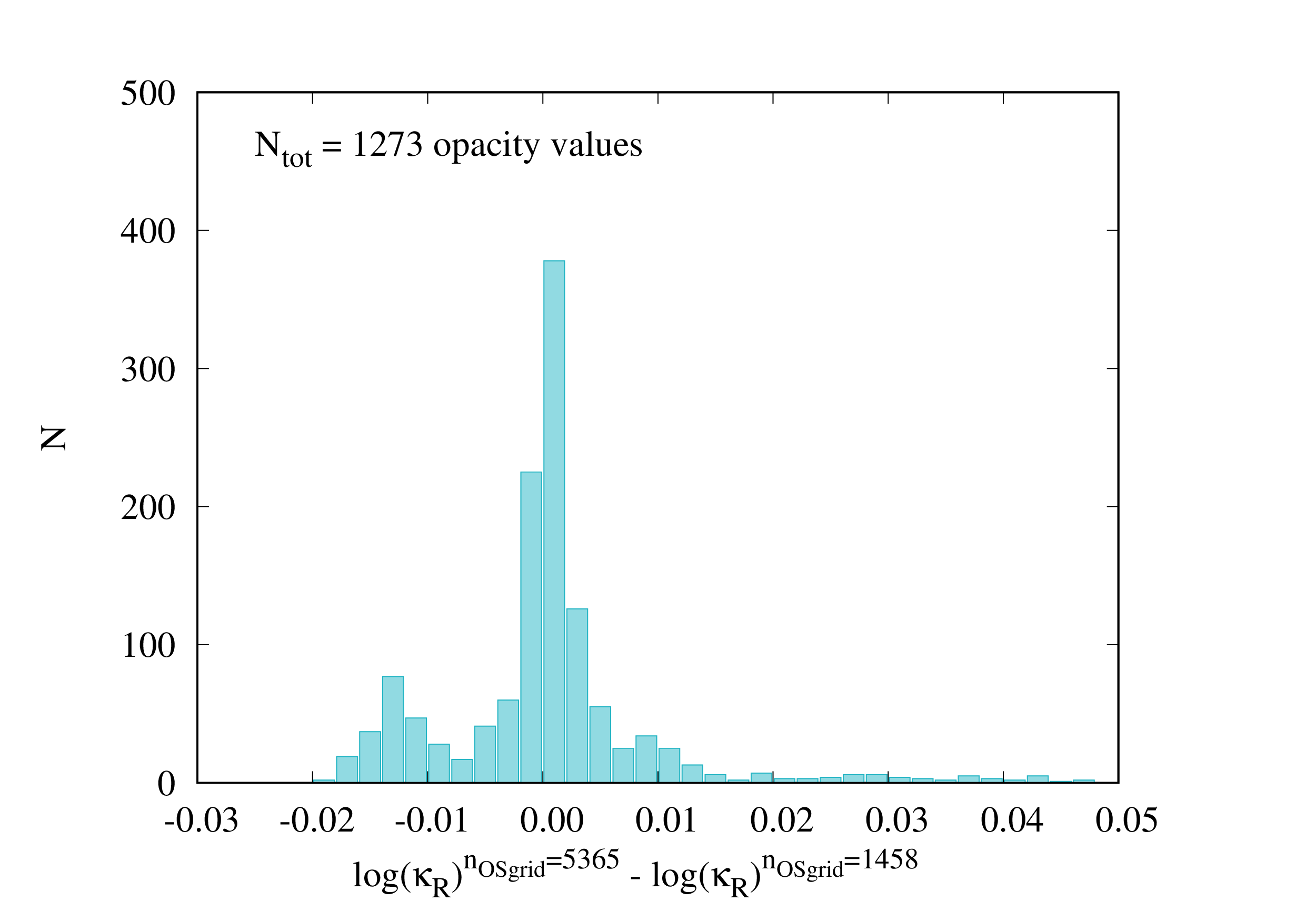}     
    \caption{Differences in Rosseland mean  opacities between the reference frequency grid with $n_{\rm OSgrid}=5365$ points, and two test cases with $n_{\rm OSgrid}=2358$ and $n_{\rm OSgrid}=1458$ (top and bottom panels, respectively).
    The chemical composition  assumes $X=0.7$, $Z=0.0165$, with scaled-solar elemental abundances according to \citet{Magg_etal_22}.
    {\em Left panels}: Maps of opacity differences across the whole $R-T$ extension of a typical table. Contour levels are distributed every $0.01$ dex in $\Delta\log(\kR)$.
    {\em Right panels}: The distribution of opacity differences with respect to the reference frequency grid, calculated over the entire sample of $\mathrm{N_{\rm tot}=1273}$ opacity values in the table. \label{fig_osmaps}}
\end{figure*}
Finally, it is worthwhile to investigate the impact of further reducing the number of frequency points below the reference grid,
with a nominal size of $n_{\rm OSgrid}=5365$.
For this purpose, we tested two additional frequency distributions with $n_{\rm OSgrid}=2358$ and $n_{\rm OSgrid}=1458$, both built using \citet{Helling_Jorgensen_98}'s optimization scheme.
Figure~\ref{fig_osmaps} illustrates the results. It is clear that reducing the frequency grid by a factor of about 2.3 or 3.7 has a limited impact, resulting in a precision loss that is mostly confined to the range of $0.01-0.02$ dex in $\log(\kR)$. At the same time, the increase in computational speed is noticeable. When using frequency grids with $n_{\rm OSgrid}=5365,\, 2358,\, 1458$ points, the CPU time required to compute the same opacity table on a laptop is 157 s, 69 s, 48 s, respectively.
We conclude that a faster performance is possible with a minimal precision loss in \kR. The next step is to assess the impact of these opacity differences on stellar models. Section~\ref{sec_evtest} addresses this critical aspect.

\paragraph{Line Broadening}
We account for line broadening due to thermal Doppler effect and non thermal-contribution of micro-turbulent velocities, by constructing a normalized broadening profile, $\phi(\nu)$, according
to the equation:
\begin{equation}
\phi(\nu)=\frac{1}{\Delta_{\nu}\sqrt{\pi}}\,e^{-\left(\frac{\nu-\nu_0}{\Delta_{\nu}}\right)^2}\,,
\end{equation}
where $\nu_0$ is the line center position in frequency, and $\Delta_{\nu}$ is the line width, computed with 
\begin{equation}
\Delta_{\nu}=\frac{\nu_{0}}{c}\sqrt{\frac{2 k_{\rm B}T}{m}+\xi^2}.
\label{eq_broadening}
\end{equation}
Here, $c$ is the speed of light, $k_{\rm B}$ is the Boltzmann constant, $m$ is the molecule's mass and $\xi$ is the micro-turbolent velocity, which is set to $2.5$~km/s. This value is compatible  with the micro-turbulent velocities inferred from stellar spectra of giant and dwarf stars \citep[e.g.,][]{Plez_etal_93, Vanture_Wallerstein_02, Mucciarelli_11}.

We emphasize that the \texttt{EXOCROSS} program employs a gaussian profile, rather than a Voigt profile. This should not have a noticeable effect on the Rosseland mean opacity, given that the many different opacity sources overlap in ways that minimize any effects of ignoring the far wings of molecular lines. In the case of planetary and brown dwarf atmospheres with little or no ionization and H appearing primarily as H$_2$, neglecting the line extended wings may be instead more significant.

\section{Discussion of the results}
\label{sec_results}
Below we will analyze the new results, comparing them with those obtained with the initial version of \texttt{\AE SOPUS} and other opacity data in the literature.

\subsection{Scaled-solar mixtures}
\label{ssec_sun}
\begin{table}\centering
\begin{threeparttable}[]
\caption{Main Solar Chemical Compositions in Literature\label{tab_sun}}
\begin{tabular}{lccccc}
\midrule
\midrule
Reference  & $(Z/X)_{\odot}$ & $Z_{\odot}$ & $(\co)_{\odot}$ & $\mathrm{(O-C)}_{\odot}$\tnote{a}
& $(\co)_{{\rm crit}}$\tnote{b}\\
\midrule
\cite{AG_1989} (AG89) & 0.02742 &0.0194 & 0.427 & 8.688 & 0.958\\
\cite{GN_93}  (GN93) & 0.02444 &0.0173 & 0.479 & 8.587 &0.952\\
\cite{GS_98} (GS98) & 0.02308 & 0.0170 & 0.490 & 8.538 & 0.947 \\
\cite{Holweger_01} (H01)\tnote{c} & 0.02094  & 0.0149 & 0.718 & 8.187 &0.937 \\
\cite{Lodders_03} (L03) &  0.01760   & 0.0132 & 0.501 & 8.388 & 0.929 \\ 
\cite{Grevesse_etal_07}  (GAS07) & 0.01653 &0.0122 & 0.537 & 8.326 & 0.929 \\
\cite{Asplund_etal_09} (AGSS09) & 0.01813 & 0.0134 & 0.549 & 8.344 & 0.934 \\
\cite{Caffau_etal_2011} (C11)\tnote{d} & 0.02070& 0.0152 & 0.575 & 8.414 & 0.938 \\
\cite{Asplund_etal_21} (AAG21) & 0.01867 & 0.0139 & 0.589 & 8.304 & 0.888 \\
\cite{Magg_etal_22} (MBS22)\tnote{e} & 0.02250 & 
0.0165 & 0.617 & 8.354 & 0.934 \\
\midrule
\end{tabular}
\tablecomments{
 For each  mixture the solar  metallicity-to-hydrogen abundance ratio $(Z/X)_{\odot}$, 
the present-day total metallicity $Z_{\odot}$ (in mass fraction), 
the ratio  (C/O)$_{\odot}$, the oxygen excess (O-C)$_{\odot}$, and (C/O)$_{{\rm crit}}$ are indicated for comparison. Carbon and oxygen abundances are expressed as number fractions.}
\begin{tablenotes}
\item[a]Following a standard notation, we define $(\mathrm{O-C})_{\odot}= \log(n_{\rm C}/n_{\rm H}-n_{\rm O}/n_{\rm H})+12$, where $n_{\rm C}$, $n_{\rm O}$, and $n_{\rm H}$ denote the number densities  of carbon, oxygen, and hydrogen in the Sun's photosphere, respectively.
\item[b]This abundance ratio is defined (C/O)$_{\rm crit}=1-n_{\rm Si}/n_{\rm O}$ \citep{FerrarottiGail06}. It marks a critical boundary for the gas molecular chemistry and opacity in the range 
$3.2~\le~\log(T)~\le~3.6$. For $\co_{\rm crit}\la \co \la 1$ the opacity enters in a narrow and deep minimum \citep[see][]{Marigo_Aringer_2009}.
\item[c] Revision of  C, N, O, Ne, Mg, Si, and Fe. All other 
elemental abundances are taken from \cite{GS_98}.
\item[d] Revision of Li, C, N, O, P, S, K, Fe, Eu, Hf, Os, Th. All other  elemental abundances are taken from \cite{GS_98}.
\item[e] Revision of all nuclides from C to Ni; Ba abundance is from \cite{Gallagher_etal_20}; Eu, Hf, Os, Th come from \cite{Caffau_etal_2011}; all other elemental abundances are taken from \cite{GS_98}.
\end{tablenotes}
\end{threeparttable}
\end{table}
In stellar models scaled-solar mixtures represent a reference choice for several applications. The chemical composition of the Sun (derived from the Sun's spectrum and/or chemical analyses of primitive meteorites) has undergone numerous revisions over the years. Table~\ref{tab_sun} lists the major solar mixtures in chronological order, beginning with the oldest \citep{AG_1989} and progressing to the most recent one \citep{Magg_etal_22}.
As can be seen, the estimated current metallicity of the Sun, \Zsun, shows a decreasing trend over time, passing from a maximum of $\Zsun = 0.0194$ \citep{AG_1989} to a minimum of around $\Zsun \simeq 0.012-0.014$ \citep{Grevesse_etal_07, Asplund_etal_09, Asplund_etal_21}.
Such decrease of \Zsun\ has entailed severe difficulties in reproducing the helioseismological constraints and the global parameters of the Sun at the present time \citep[e.g.,][]{Serenelli_etal_09}.
\cite{Magg_etal_22} most recent revision of the standard composition of the Sun indicates a new increase in the present-day metallicity (primarily due to greater abundances of oxygen and carbon) up to $\Zsun = 0.0165$, with a significant improvement in the standard solar model's ability to reproduce observational data. The calibration of the solar model with our \texttt{PARSEC} code \citep{Bressan_etal_12}, based on the new  \texttt{\AE SOPUS\,2.0} opacities, is currently in progress.

\begin{figure*}[h!]
    \centering
    \includegraphics[width=0.60\textwidth]{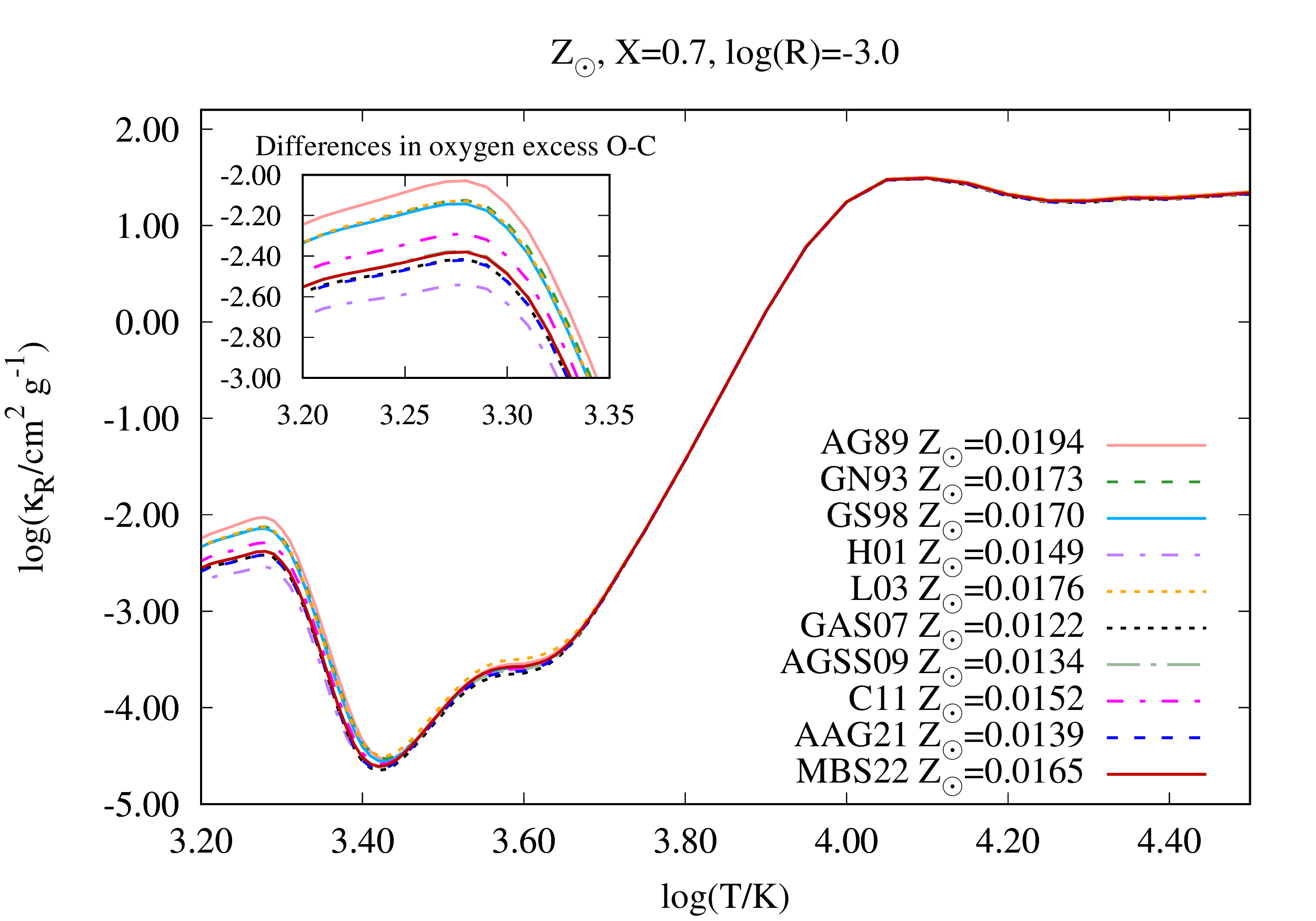}
\caption{\texttt{\AE SOPUS\,2.0} Rosseland mean opacity as a function of temperature, with $\log(R)=-3$. The adopted chemical composition is defined by $X = 0.7$, and $Z=\Zsun$, according to various
compilations of the solar mixture (see Table~\ref{tab_sun}). 
As highlighted in the inset, the largest differences occur at $\log(T/{\rm K})\la 3.4$, where water absorption dominates the opacity.\label{fig_ksol_var}}
\end{figure*}

Figure~\ref{fig_ksol_var} displays \kR\ as a function of temperature for scaled-solar mixtures according to all solar compositions in Table~\ref{tab_sun}. We note that the present-day solar photospheric carbon-to-oxygen ratio, \co,  varies between sources, from a minimum of $\co\simeq0.427$ in AG89, to larger values such as $\co\simeq0.718$ of H01, and more recently, $\co\simeq0.617$ in MBS22.
These variations may produce a significant impact on molecular chemistry and opacity mostly for $\log(T/{\rm K}) \la 3.4$.
More specifically, at these temperatures, \kR\ is dominated by the opacity bump caused by the H$_2$O molecule, the magnitude of which is extremely sensitive to the excess of oxygen atoms over those of carbon, $\mathrm{O-C}$.
This parameter, which is also listed in Table~\ref{tab_sun}, represents the amount of oxygen that is not locked in the extremely stable CO molecule and is available for the formation of important O-bearing  absorbing  species, such as H$_2$O, AlO, VO, TiO, ZrO.
As a result, the larger the $(\mathrm{O-C})_{\odot}$, the higher the \kR\ because more oxygen is free to be trapped in H$_2$O.

This is evident when we compare the opacity curves of MBS22 and GS98 in Figure~\ref{fig_ksol_var}.  Despite having a similar solar metallicity ($Z_{\odot} \simeq 0.0165-0.0170$),   the opacity peak due to water at $T\simeq2000$ K in GS98 is higher due to a greater oxygen excess, $(\ominusc)_{\odot} \simeq 8.538$ compared to $(\ominusc)_{\odot} \simeq 8.354$ in MBS22. At the same time, despite significant differences in $Z_{\odot}$, AGSS09 and MBS22 exhibit an almost identical opacity profile at $1600 \la T \la 2500$ K, owing to a comparable $(\ominusc)_{\odot} \simeq 8.34-8.35$.


\subsection{Comparison of \AE SOPUS\,1.0, 2.0 and Ferguson et al. (2005)}
\begin{figure*}[h!]
    \centering
     \includegraphics[width=0.48\textwidth]{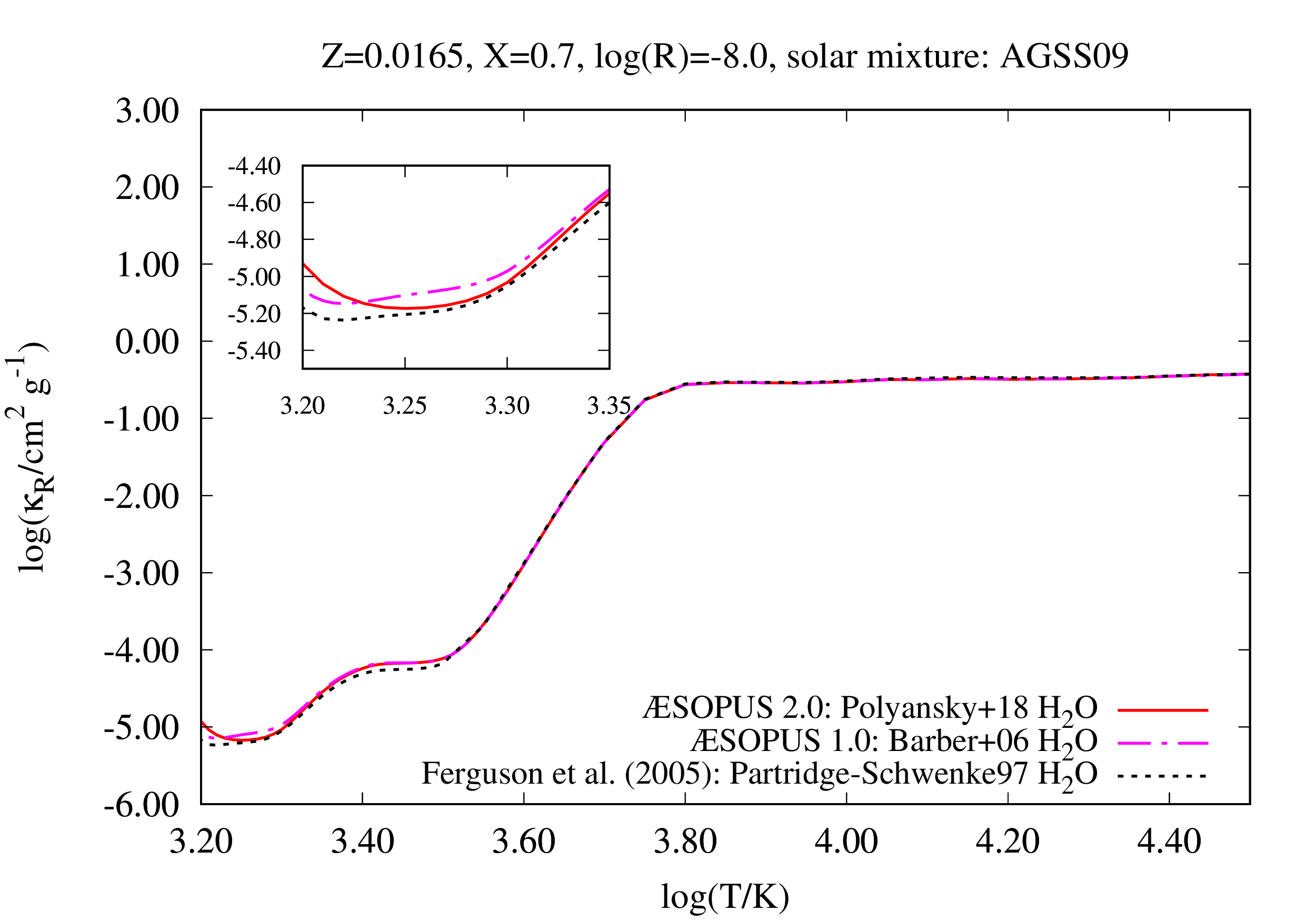}
     \includegraphics[width=0.48\textwidth]{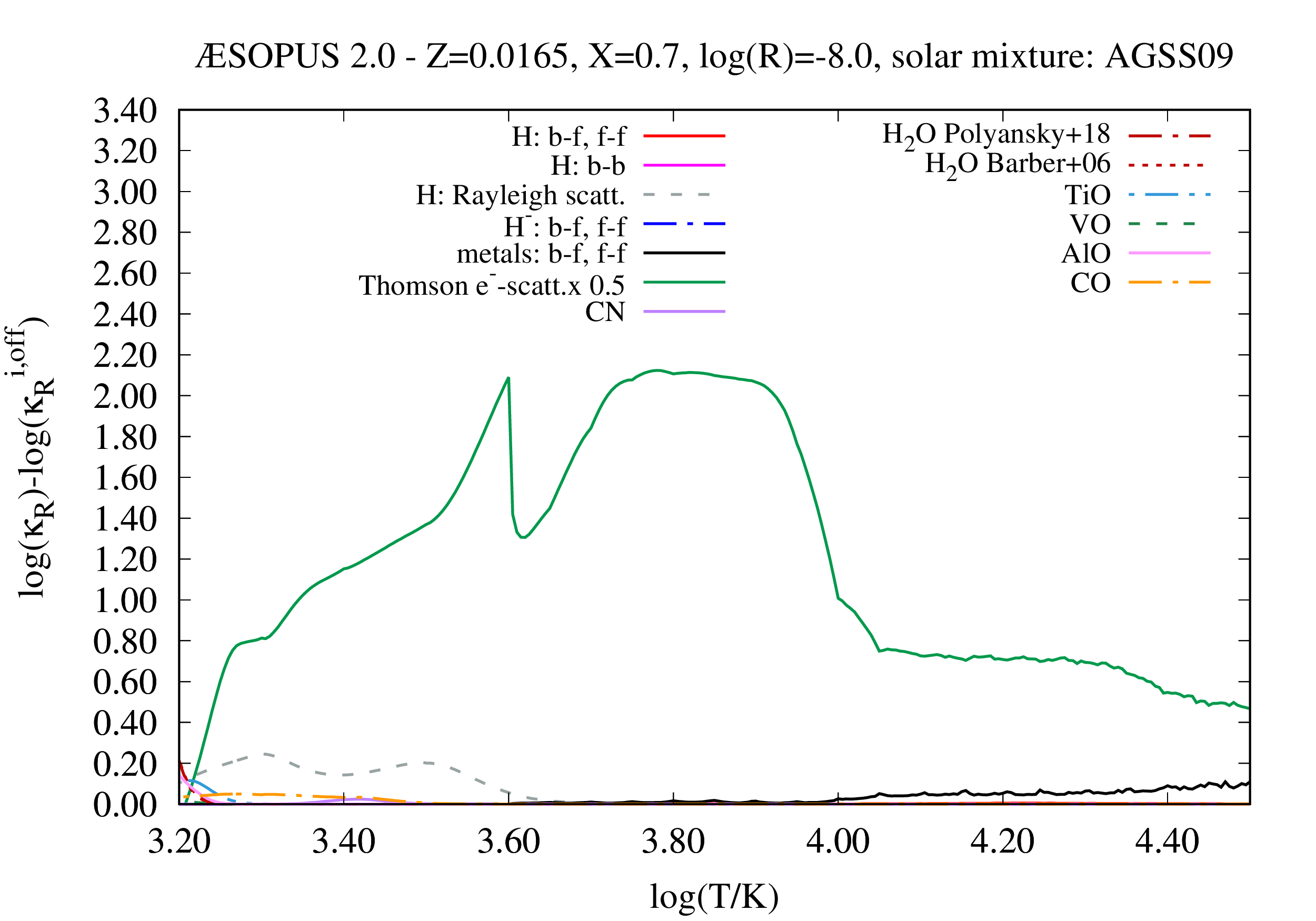}
    \includegraphics[width=0.48\textwidth]{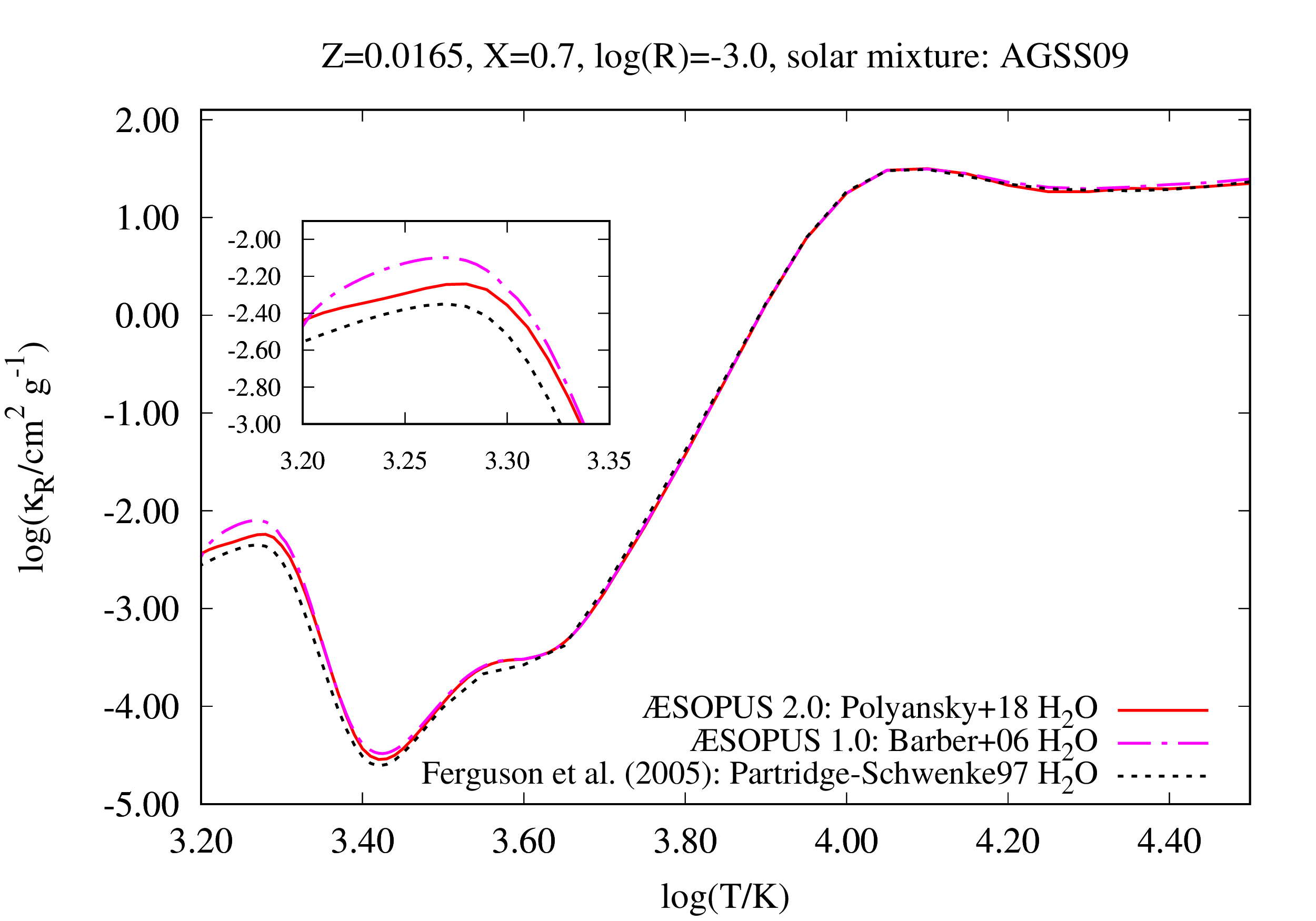}
     \includegraphics[width=0.48\textwidth]{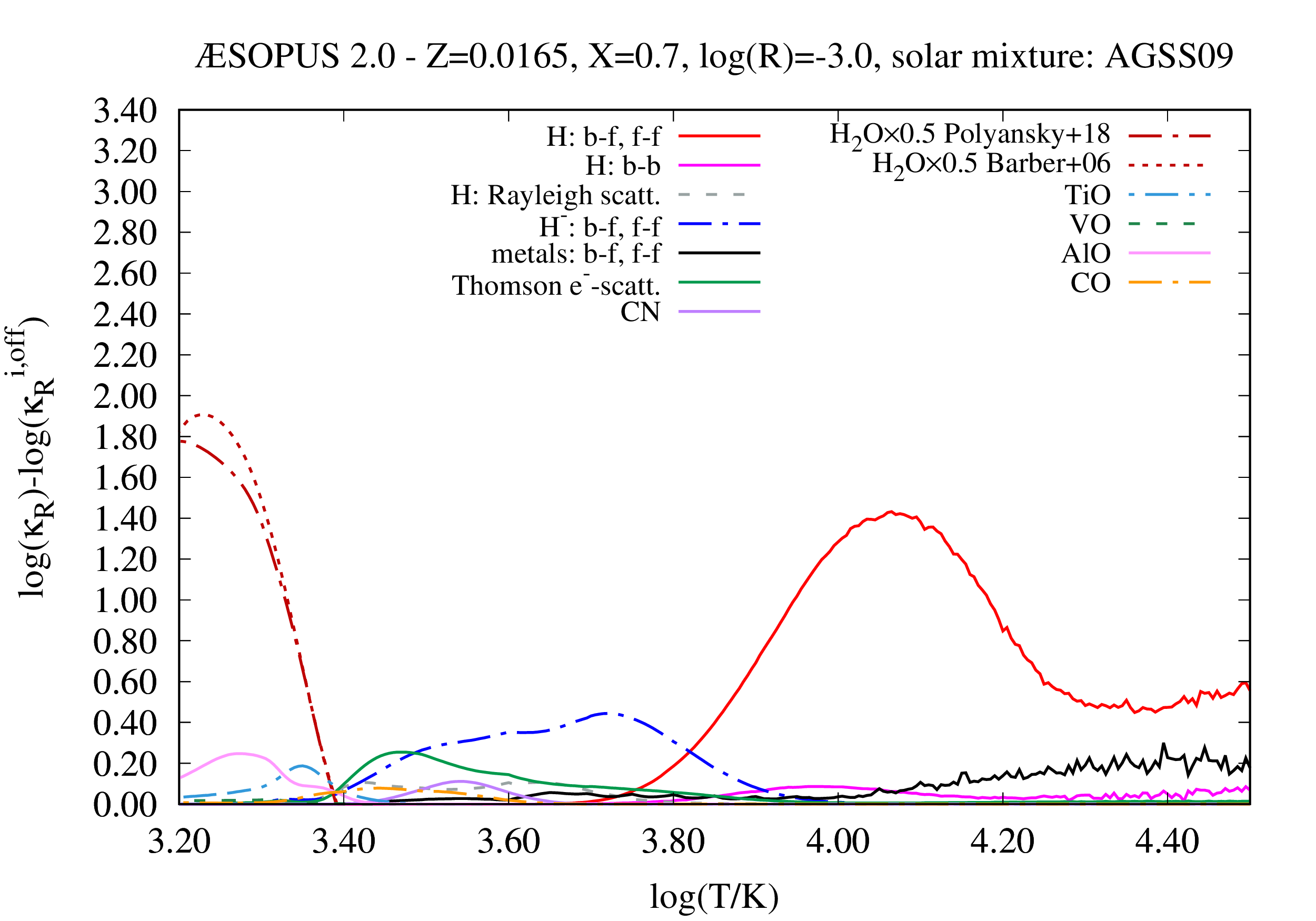}
    \includegraphics[width=0.48\textwidth]{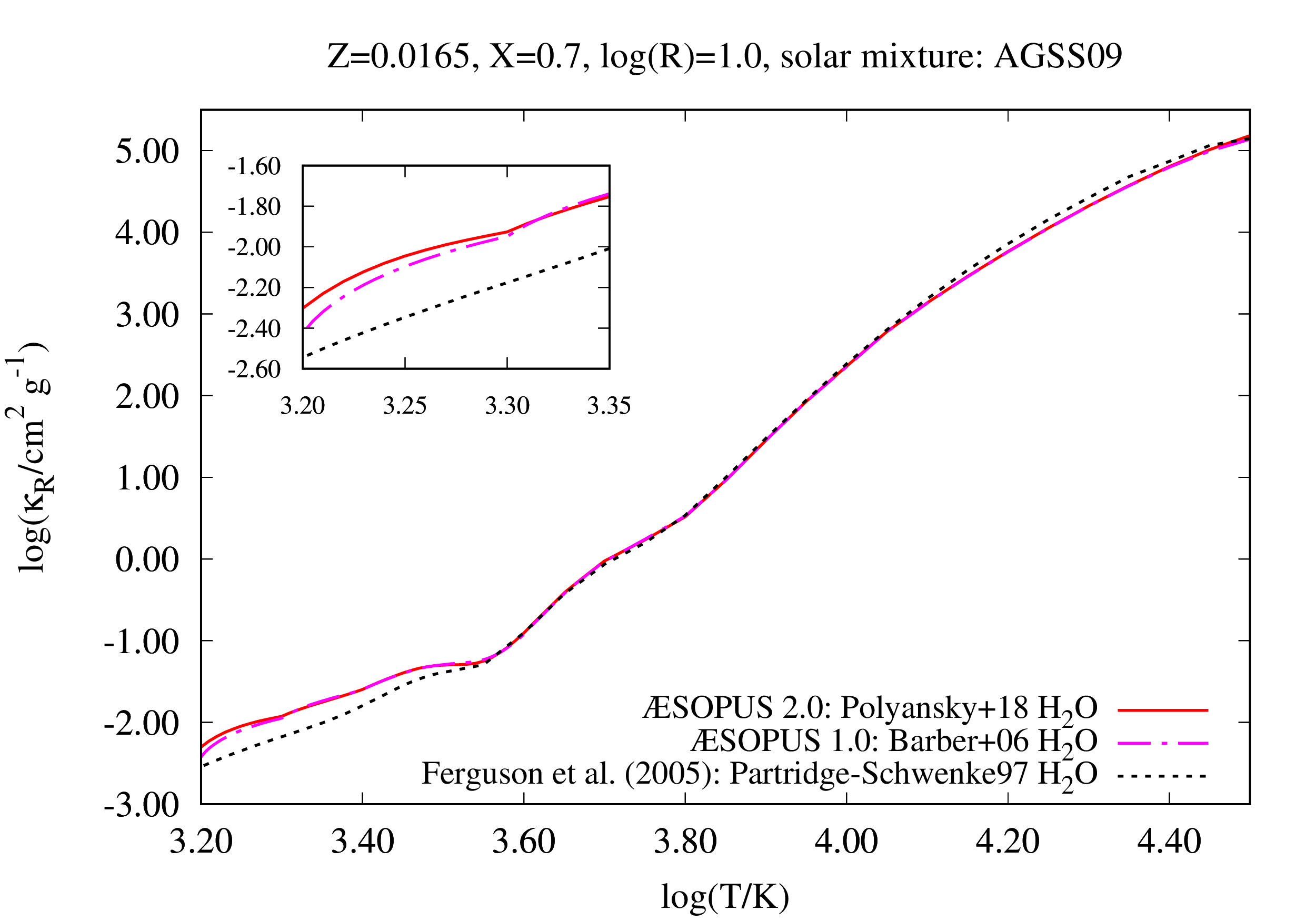}
     \includegraphics[width=0.48\textwidth]{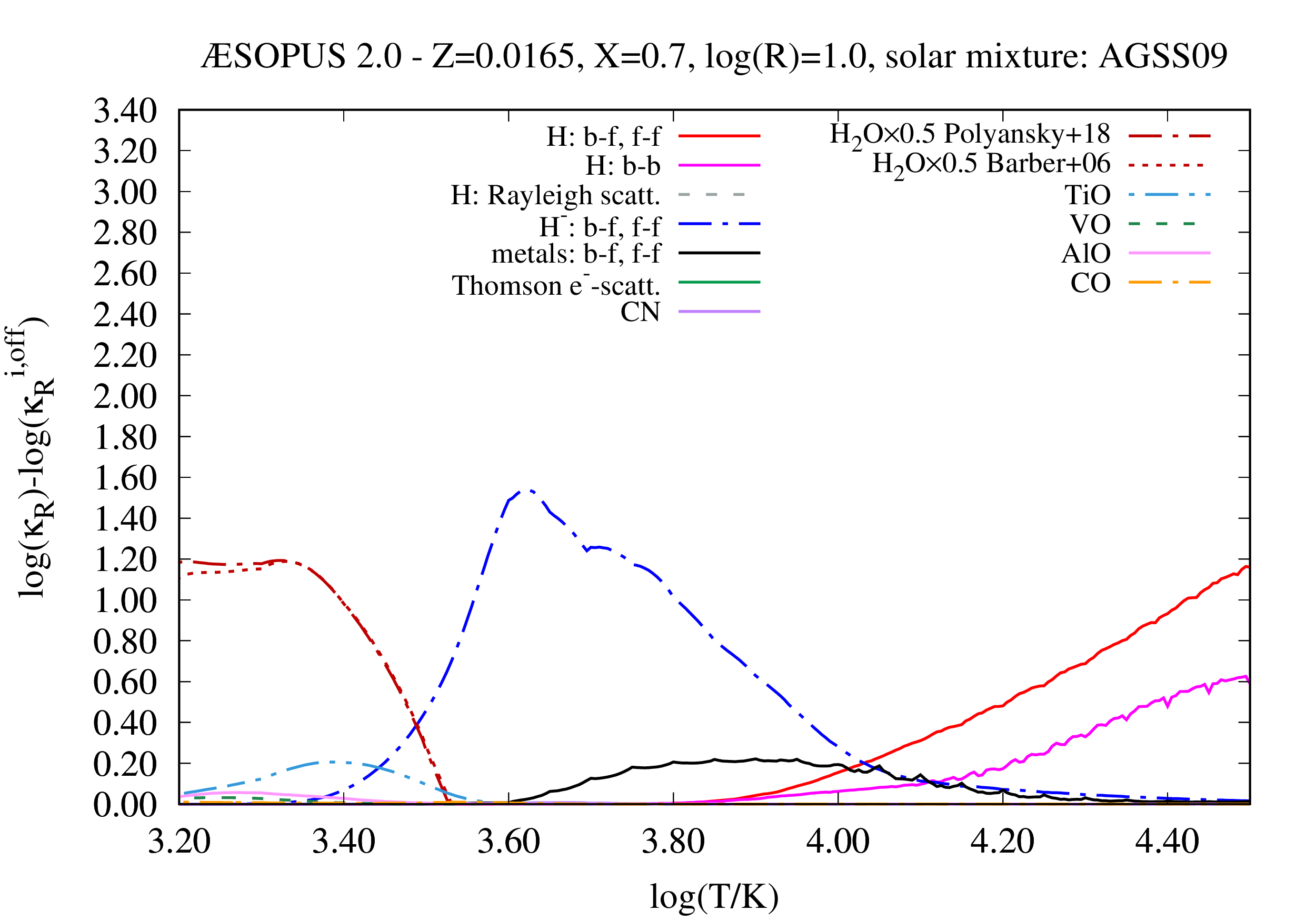}
\caption{Rosseland mean opacity and major opacity sources as a function of temperature $T$ and for three values of $R$.
The adopted composition
is assumed to have  $Z=0.0165$, $X=0.7$  and the metal abundances scaled to solar, following \citet{Asplund_etal_09}.
{\em Left panels}: As indicated in the legends, the three curves correspond to different opacity calculations. The insets zoom in the most pronounced opacity variations for temperatures below about $2240$ K. Except for the lowest density case (top panels with $\log(R)=-8$), where Thomson e$^-$ scattering dominates the opacity, the differences are primarily due to different water absorption line lists and the line broadening scheme used.
{\em Right panels}: Contributions to the total Rosseland mean opacity from various opacity sources, as derived from \texttt{AESOPUS 2.0} calculations shown in the left panel. Each curve corresponds to $\log(\kR)-\log(\kR^{i, {\rm off}})$, where $\kR$ is the full opacity including all opacity sources considered here, and $\kR^{i, {\rm off}}$ is the reduced opacity obtained by excluding the specific absorbing species. 
See the text for more details. \label{fig_ksol}}
\end{figure*}

Figure~\ref{fig_ksol} will help us appreciate the differences brought about by our update to the opacity sources. For this purpose, in the left panels, we compare three sets of opacity calculations: the current implementation (\texttt{\AE SOPUS\,2.0}), the first version \citep[][\texttt{\AE SOPUS\,1.0}]{Marigo_Aringer_2009}, and  \cite[][hereafter also F05]{Ferguson_etal_05}. 
The reference solar mixture is from \cite{Asplund_etal_09}.
\begin{figure*}[h!]
    \centering
    \includegraphics[width=0.48\textwidth]{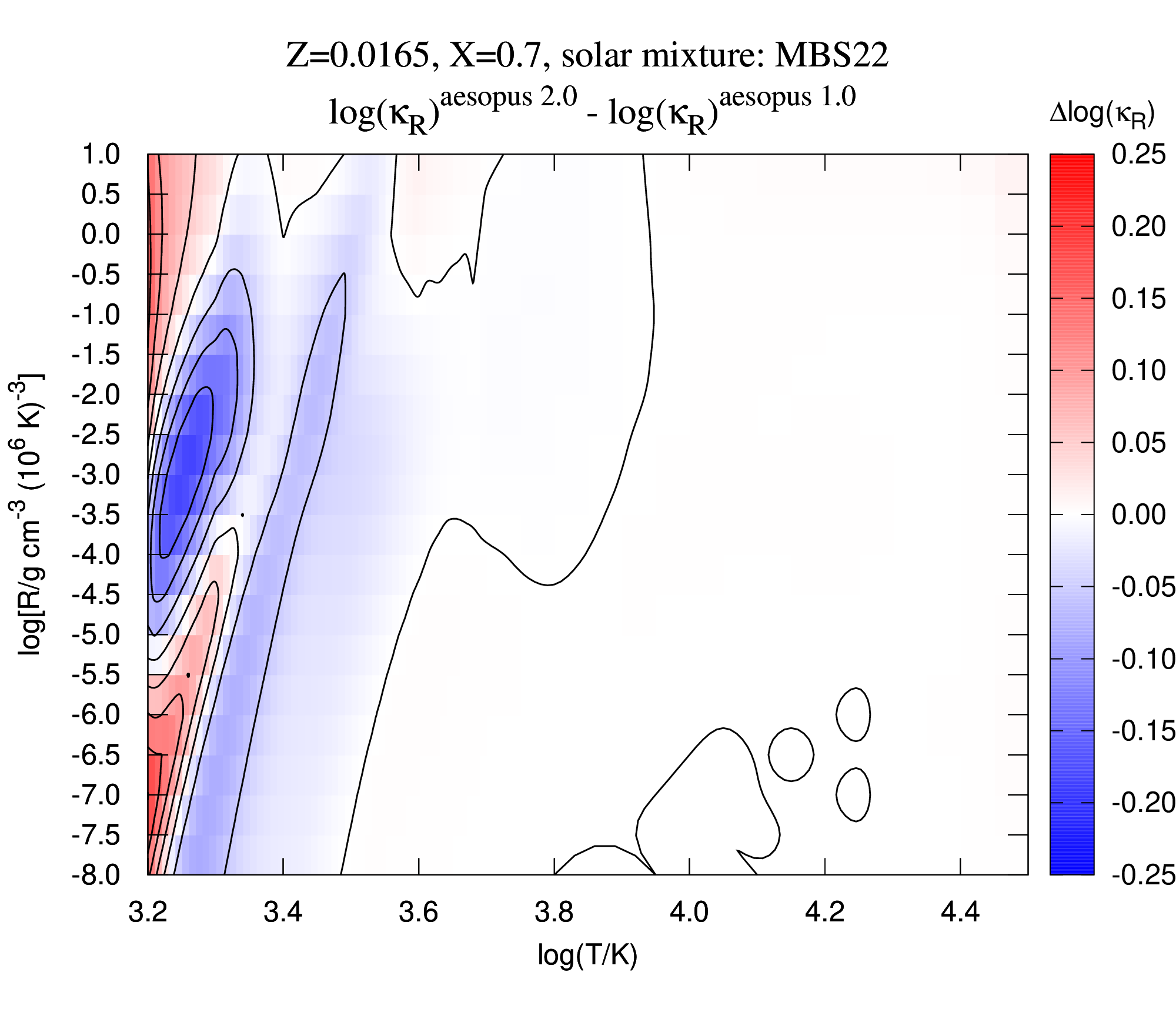}
    \includegraphics[width=0.48\textwidth]{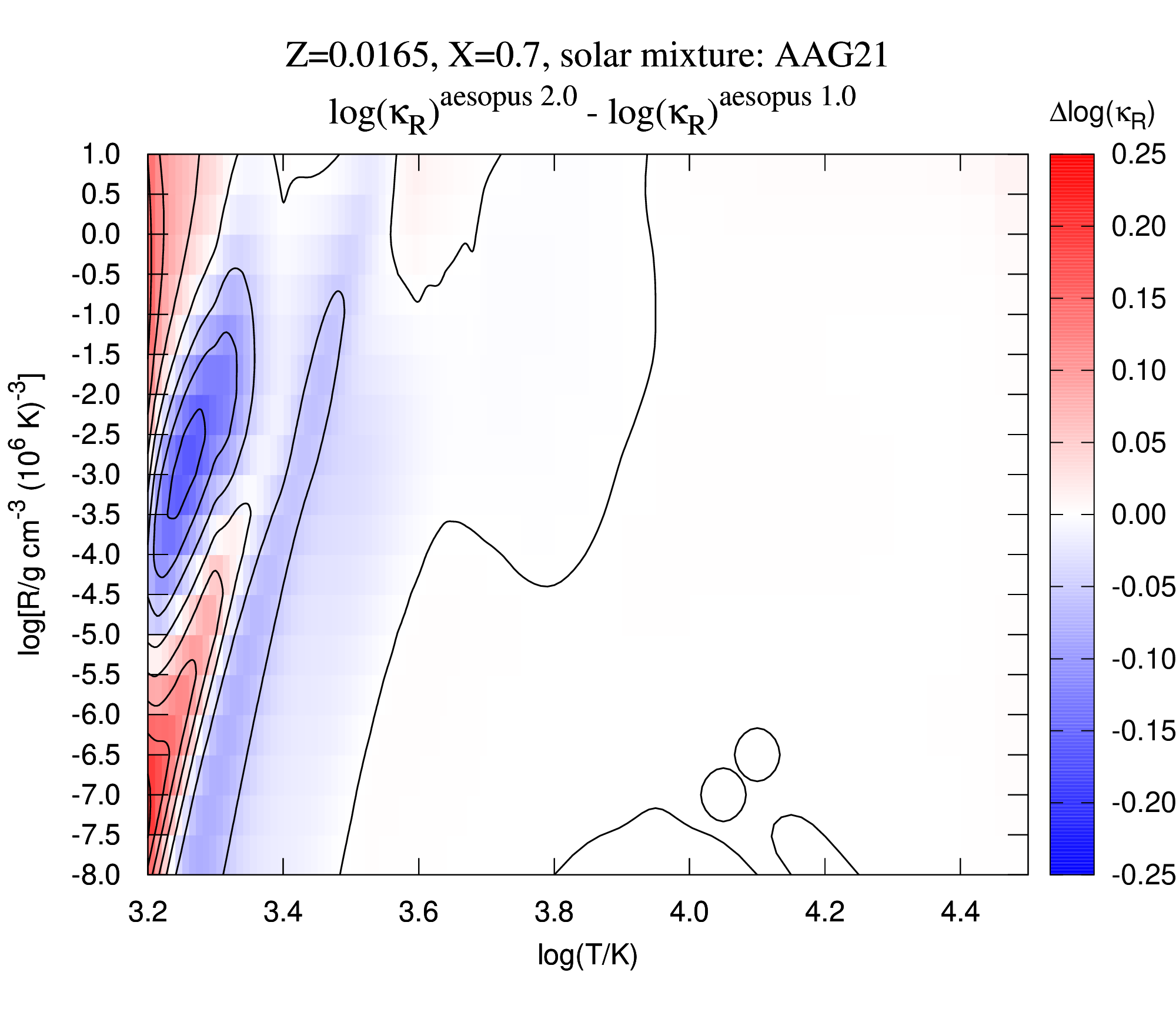}
    \includegraphics[width=0.48\textwidth]{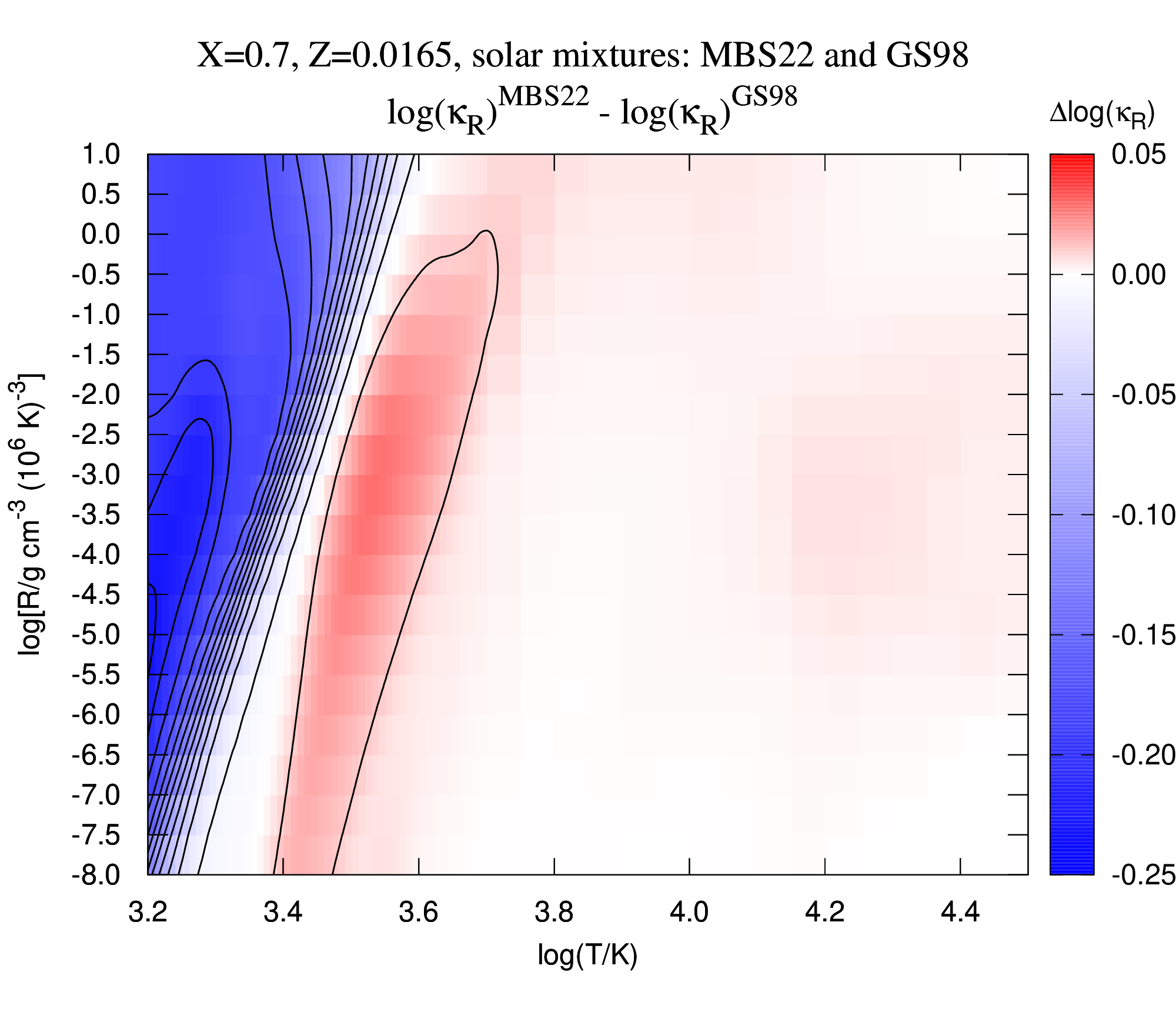} 
    \includegraphics[width=0.48\textwidth]{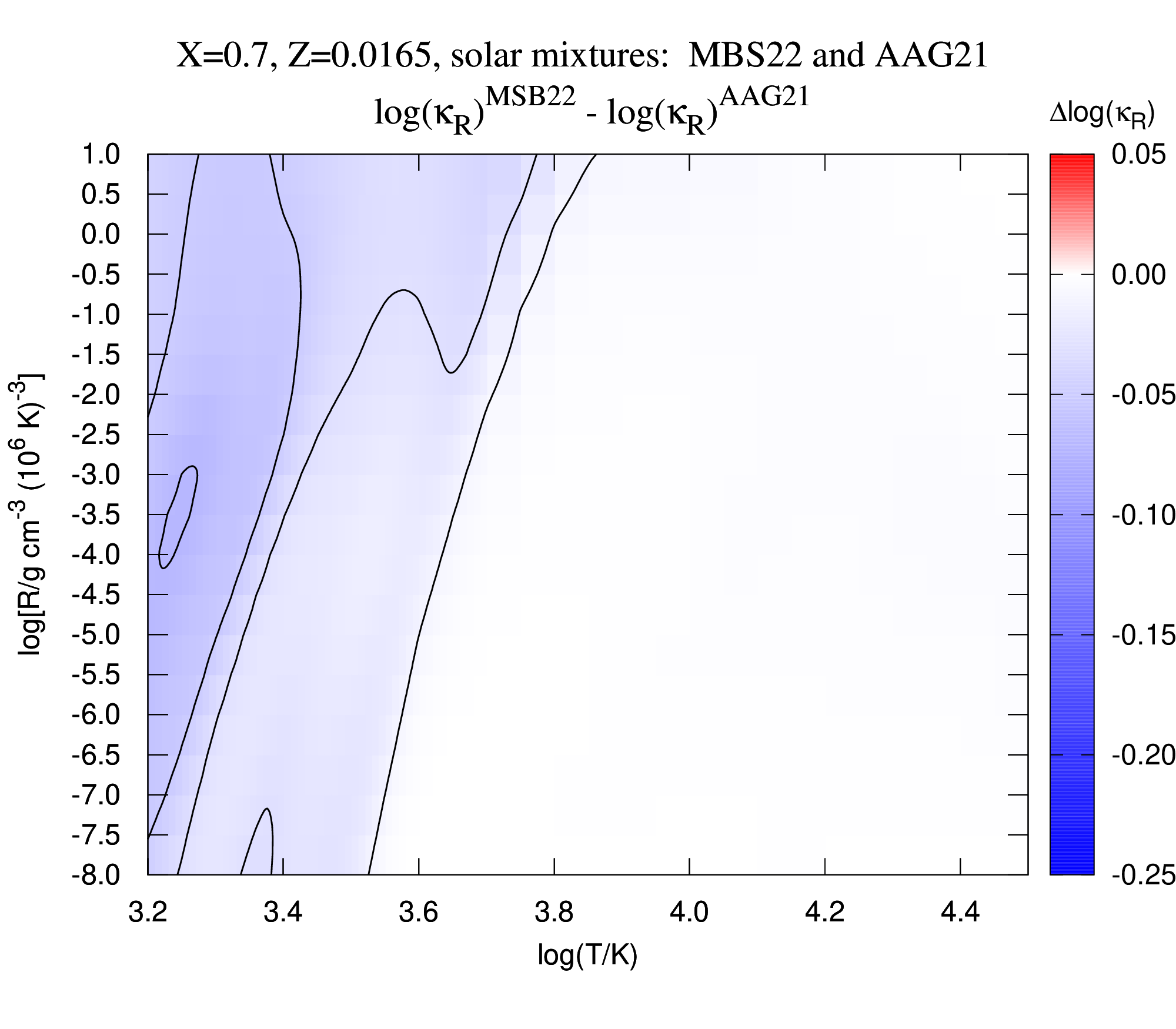}    
    \caption{Comparison of Rosseland mean opacities  across the entire extension of a typical table, assuming $X=0.7$ and $Z=0.0165$.
    {\em Top panels}: Differences between \texttt{\AE SOPUS\,2.0} and \texttt{\AE SOPUS\,1.0}, mainly due to the updates and expansion of molecular absorption. Solar compositions from
    \citet{Magg_etal_22} (left) and \citet{Asplund_etal_21} (right) are assumed. Contour levels map $0.05$ dex difference in $\log(\kR)$.
    {\em Bottom panels}: Differences caused by the choice of solar mixture. Comparison is made between GS98 and MBS22 (left), or AAG21 and MBS22 (right). Version \texttt{\AE SOPUS\,2.0} is adopted. The grid of contour levels is distributed every $0.02$ dex difference in $\log(\kR)$. \label{fig_map_diff}}
\end{figure*}
\begin{figure*}[h!]
    \centering
    \includegraphics[width=0.48\textwidth]{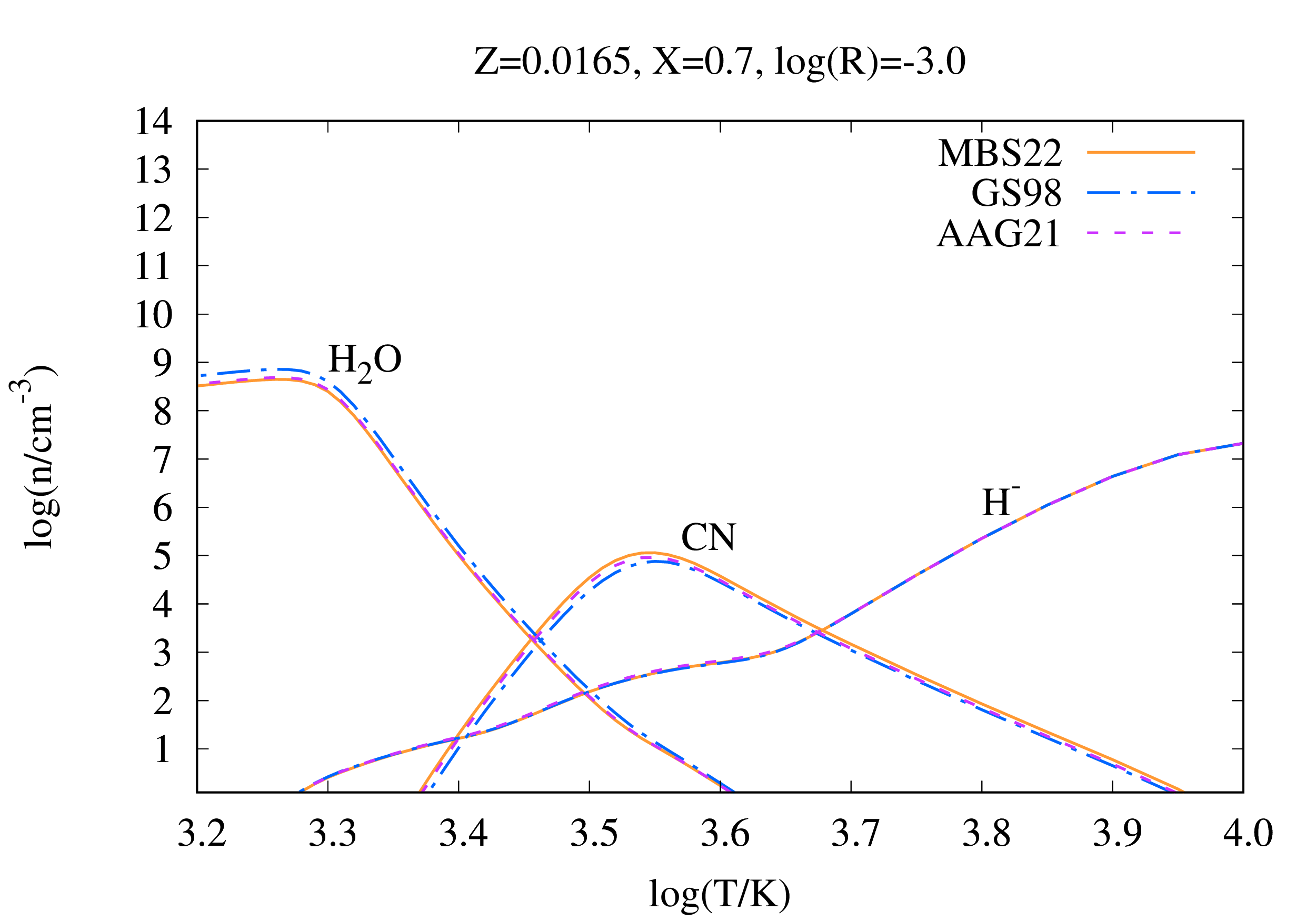}
     \includegraphics[width=0.48\textwidth]{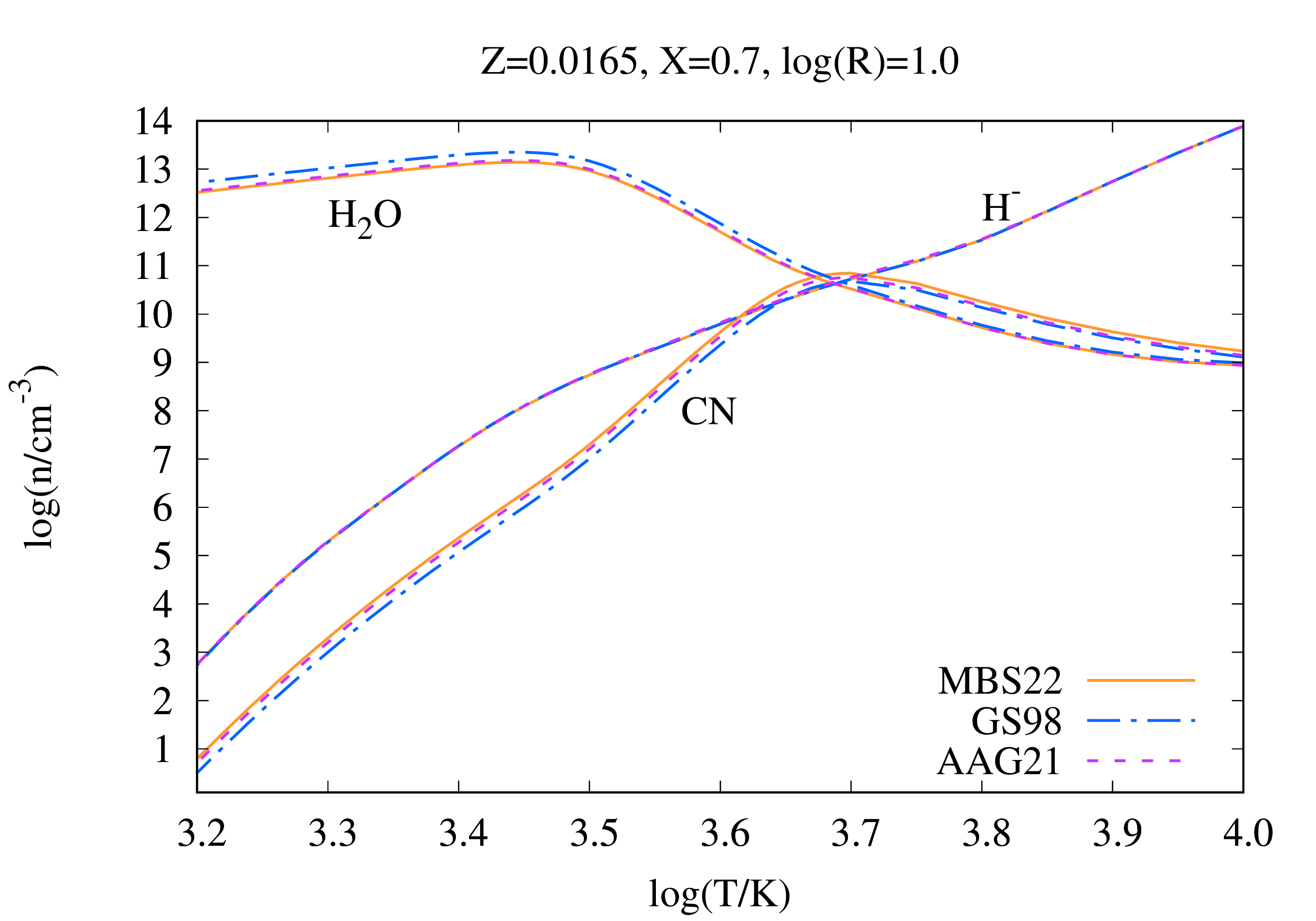}
\caption{Number densities of H$_2$O, CN, and H$^-$ as a function of temperature for two $R$ parameter values. \texttt{\AE SOPUS\,2.0} predictions are shown for three solar mixtures, namely GS98, AAG21, MBS22. \label{fig_nmol}}
\end{figure*}

It is also useful to refer to the right panels, which display the temperature windows where the main opacity sources make a significant contribution.  For each opacity source $i$ we compute the quantity
$\log(\kR)-\log(\kR^{i, {\rm off}})$, where $\kR$ is the total Rosseland mean opacity including all opacity sources considered here, and $\kR^{i, {\rm off}}$ is the reduced opacity obtained by ignoring the species $i$. This specific notation allows us to highlight the temperature domains where different opacity sources contribute the most.

The case with $\log(R)=-8$ (top panels) corresponds to a very low density regime in which the opacity is almost entirely dominated by Thomson electron scattering, with some contribution from H Rayleigh scattering for $\log(T)<3.6$. Molecular absorption plays a minor role. This explains the very small differences between the various sets of opacity calculations.
In the case with $\log(R)=-3$ (middle panels) differences start to appear for $\log(T/{\rm K})\la 3.6$. They remain  moderate in the range $3.4 \la \log(T/{\rm K}) \la 3.6$, which is likely due to the revision of the opacity of the CN molecule. 
In terms of CN, this work employs the line list of \cite{CN_10.1093/mnras/stab1551}, \cite{Marigo_Aringer_2009} uses the data from  \cite{Jorgensen_97}, and \cite{Ferguson_etal_05} adopts the line list of \cite{JorgensenLarsson_90}.

We find that the revision of H$^-$ photodetachment cross section \citep{McLaughlin_2017} produces very small differences in the resulting \kR\  when compared to previous predictions \citep{Wishart_79}, both at intermediate ($\log(R)=-3$, middle panels) and high densities ($\log(R)=1$, bottom panels).
This fact can be explained as follows. The  resonances of H$^-$ photodetachment  are located at $h\nu > 10$ eV, which correspond to normalized photon energies $u > u_{\rm res}$, with the exact value of $u_{\rm res}$ depending on the gas temperature.
It is easy to see that at the temperatures ($3000 \la T/{\rm K} \la 8000$) where H$^-$ contributes substantially to \kR, $u_{\rm res}$ varies in the range
$38.7 \ga u_{\rm res} \ga 14.5$, where the weighting function $F_{\rm R}$ of the Rosseland mean opacity is very small or close to zero (see Figure~\ref{fig_fr} and Equation~(\ref{eq_krossu})). As a result, the resonances have only a minor effect on the integral that defines \kR.

At lower temperatures, in the interval $3.2 \la \log(T/{\rm K}) \la 3.35$, and examining the case with $\log(R)=-3$ (middle panels), 
Figure~\ref{fig_ksol}  clearly shows that \texttt{\AE SOPUS\,2.0} revised opacity lies somewhere in between \texttt{\AE SOPUS\,1.0}  and F05.
In the latter temperature range  $\kappa_{\rm R}$ is dominated by water molecular absorption, and therefore it is affected by the adopted H$_2$O line list. We recall that F05 adopts \cite{Partridge_Schwenke_97}, \cite{Marigo_Aringer_2009} uses the \texttt{BT2} transitions from \cite{Barber_etal_06}, whereas in this work we employ the \texttt{POKAZATEL} line list from \citet[][see Figure~\ref{fig_h2o_tio}]{H2O_10.1093/mnras/sty1877}. 
Different line broadening schemes are likely to cause additional discrepancies. F05 uses a thermal Doppler profile; we do the same but also include the effect of micro-turbulence velocity (see Equation~(\ref{eq_broadening})).



It is worthwhile to compare the differences in Rosseland mean opacity caused by changes in the input data (e.g, molecular line lists, line profiles, other opacity sources). Figure~\ref{fig_map_diff} illustrates a few examples. Looking at the top panels, we can see that the opacity changes from \texttt{\AE SOPUS\,1.0} to \texttt{\AE SOPUS\,2.0} are  distinguishable but not dramatic. The water opacity bump, in particular, is reduced by up to $0.25$ dex. As expected, these changes are largely independent of the solar composition: using the MBS22 (top-left panel) or AAG21 (top-right panel) produces nearly identical maps of opacity difference.

\subsection{Changes in Solar Mixture Effects}
Finally, we investigate the main opacity differences caused by different solar mixture options, while keeping the input opacity data constant. The bottom panels of Figure~\ref{fig_map_diff} show the results of a few tests.
Three alternatives are being considered: GS98, AAG21, and MBS22, all with the same metallicity $Z=0.0165$ and hydrogen abundance $X=0.7$. 
The largest differences appear at $\log(T/\rm{K}) < 3.8$, where molecular absorption becomes significant and is influenced by the relative distribution of elemental abundances.
To aid the discussion, Figure~\ref{fig_nmol} depicts the number densities of three species (H$_2$O, CN, H$^-$), which have been shown to significantly contribute to the Rosseland mean opacity at these temperatures.

Let us first focus on the role of H$^-$, the abundance of which is critically dependent on the availability of free electrons provided by low-ionization potential atoms, particularly  Mg, Si, Fe (see Figure 22 of \citealt{Marigo_Aringer_2009}). The concentration of H$^-$ in the three solar mixtures is nearly identical, with minor differences. The reason for this is that the total abundance of the major electron donors (Mg, Si, and Fe) varies little in the three cases, resulting in essentially the same opacity contribution from H$^-$, for  both $\log(R)=-3.0$ and $\log(R)=1.0$.

Another source of opacity is the absorption of CN, the concentration of which is sensitive to C and N abundances, as well as to the O excess, $(\mathrm{O-C})$. Figure~\ref{fig_nmol} shows that at any temperature, the CN abundance increases as we move along the sequence GS98, AAG21, and MBS22. How can we explain these findings? Clearly, the chemistry of CN is the result of solving the equation of state by integrating a complex system of differential equations. However, we can glean an insightful answer from the work of \cite{Scalo_74}, who developed a simplified, but valid, analytic treatment of CN chemistry and demonstrated that in stars with $\co < 1$, CN abundance scales roughly as
\begin{equation}
\mathrm{CN \propto \frac{\sqrt{N}\,C}{O-C}\equiv \tilde{A}_{CN}}   
\end{equation}
We verified that $\mathrm{\log(\tilde{A}_{CN,GS98})=3.942,\,\log(\tilde{A}_{CN,AAG21})=4.071, \, \log(\tilde{A}_{CN,MBS22})=4.196}$ using the solar abundances of the three solar compositions, which is consistent with the increasing order of the CN abundances from the full \texttt{\AE SOPUS} computation.

At lower temperatures $\log(T/{\rm K}) \la 3.4$, the H$_2$O bump contributes the most to the opacity, the abundance of which is primarily determined by the oxygen excess over carbon, $(\mathrm{O-C})_{\odot}$ (see discussion in Section~\ref{ssec_sun}). The highest H$_2$O abundance, as expected, corresponds to GS98, which has the highest $(\mathrm{O-C})_{\odot}$, whereas MBS22 and AAG21 have similar H$_2$O concentrations reflecting close  oxygen excess values (see Table~\ref{tab_sun}).

In light of these chemistry arguments, we now have the proper tools to interpret the opacity difference maps shown in the bottom panels of Figure~\ref{fig_map_diff}.
We notice the greatest opacity differences are observed between GS98 and MBS22 (left panel).
On the one hand, the higher CN abundance of MBS22 compared to GS98, may contribute to an increase in opacity by about the same amount (red region). On the other hand, the smaller $(\mathrm{O-C})_{\odot}$ of MBS22 compared to GS98 (by about 0.18 dex; see Table~\ref{tab_sun}) explains the lower MBS22 opacity where H$_2$O prevails (blue region).
When AAG21 and MBS22 are compared (right panel), the opacity differences are smaller and do not exceed 0.1 dex. The Rosseland mean opacities for MBS22 solar mixture are only slightly lower than those for AAG21 solar mixture (blue region),  reflecting the similar concentrations of H$^-$, CN, and H$_2$O.

\section{Preliminary Evolutionary Tests}
\label{sec_evtest}
While a detailed analysis of the impact of low-temperature opacities on stellar structure and evolution is beyond the scope of this paper, we discuss two illustrative cases here, namely the predicted location in the H-R diagram of the Hayashi tracks drawn by low-mass stars as they evolve through the red giant branch (RGB) and thermally-pulsing AGB (TP-AGB) phases. We used the \texttt{COLIBRI} code to perform numerical integrations of a complete envelope model that extends from the atmosphere down to the surface of the degenerate core \citep{colibri}. The \texttt{COLIBRI} code is an appropriate tool for our preliminary tests because it fully incorporates \texttt{\AE SOPUS} as a subroutine for both equation of state and opacity. The mixing-length parameter is set to $\alpha_{\rm MLT}= 1.74$. The procedure is fully described in \cite{colibri}. 
In this way, we can investigate the differences in effective temperature, $T_{\rm eff}$, caused by using \texttt{\AE SOPUS\,1.0} or \texttt{\AE SOPUS\,2.0} opacities. 
\begin{figure*}[h!]
    \centering
    \includegraphics[width=0.48\textwidth]{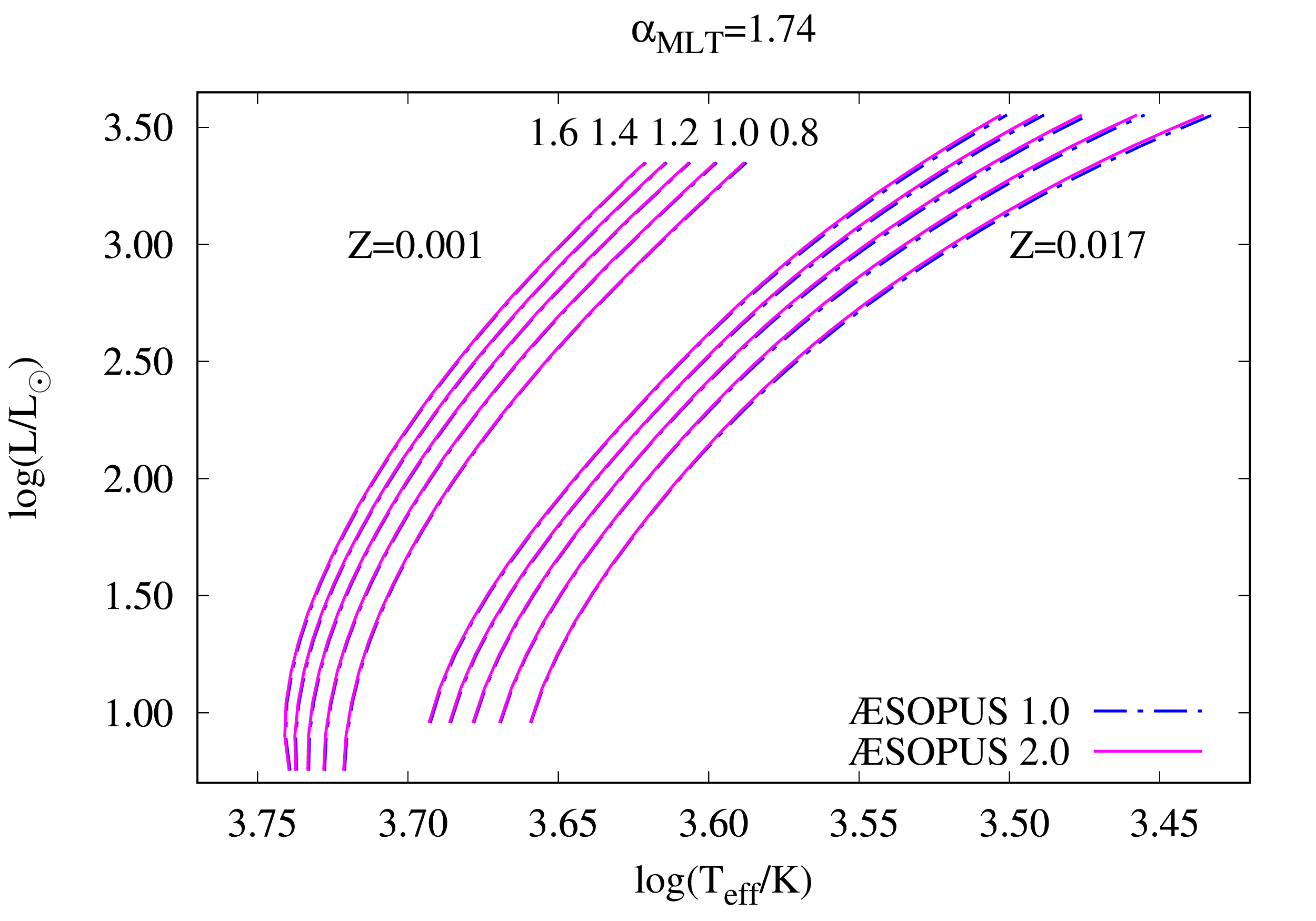}
    \includegraphics[width=0.48\textwidth]{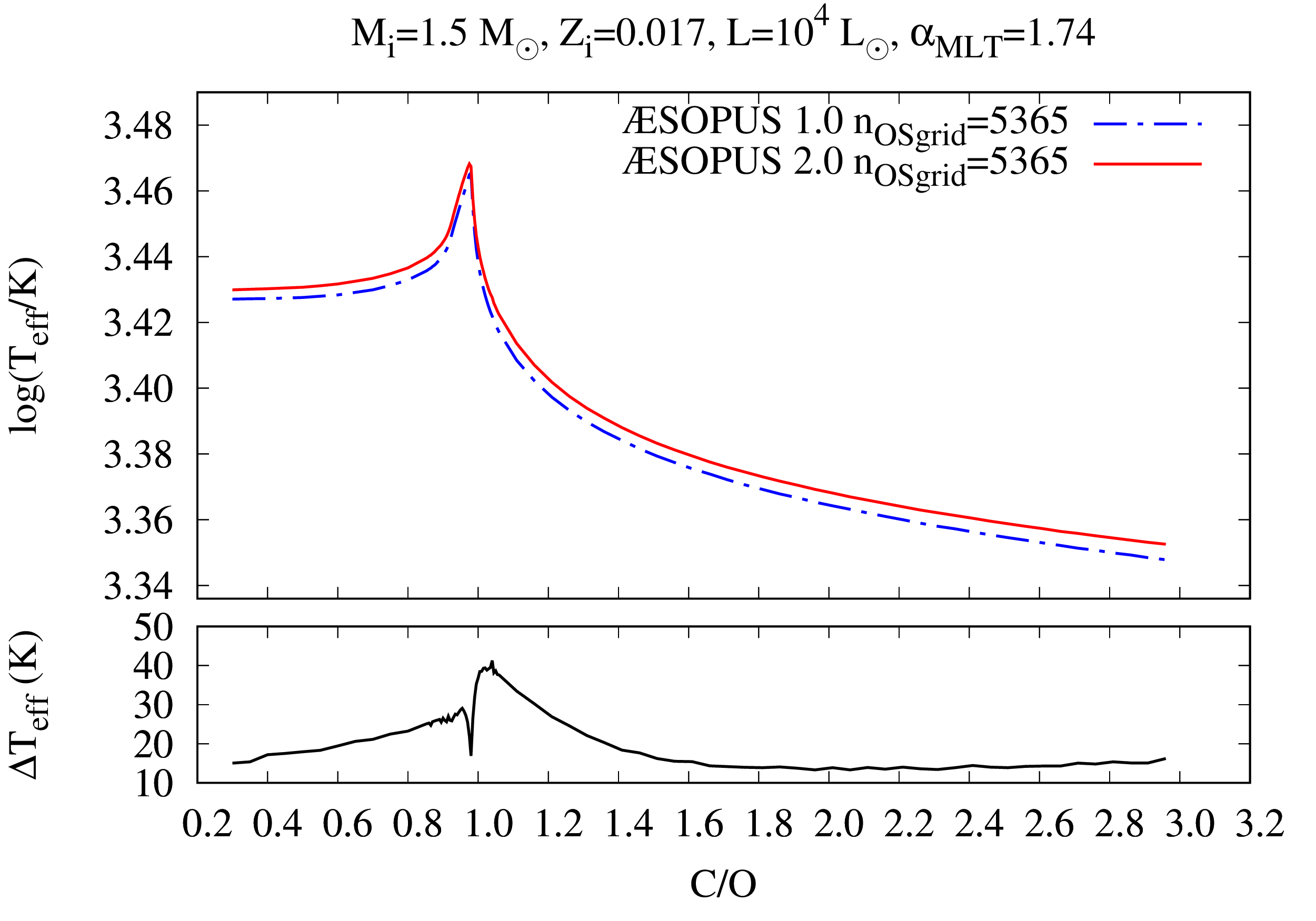}
\caption{Differences in effective temperature caused by using either \texttt{\AE SOPUS\,1.0} (blue lines) or \texttt{\AE SOPUS\,2.0} (magenta lines) opacites in red giant models. The reference solar composition is MBS22.
{\em Left panel}: RGB tracks predicted on the H-R diagram with $Z = 0.001,\,0.017$ and $M_{\rm i}$ ranging from $0.8\, \Msun$ to $1.6\, \Msun$ (with a mass step of $0.2\,\Msun$).
{\em Right panel}: Predicted effective temperature as a function of increasing photospheric C/O in a TP-AGB star (top), and difference $\Delta T_{\rm eff}$, between the two \texttt{\AE SOPUS} versions (bottom). The selected stellar parameters and frequency grid size are indicated.  See text for more details. \label{fig_hayashi}}
\end{figure*}
The left panel of Figure~\ref{fig_hayashi} depicts a series of RGB tracks with varying initial masses and two metallicity values $Z=0.001,\,0.017$. 
At given $Z$ all tracks have the  same chemical composition, extracted from \texttt{PARSEC} RGB models with $M_{\rm i}=1.0\,\Msun$, after the first dredge-up.
The luminosity is calculated using \cite{Boothroyd_Sackmann_88} core-mass luminosity relation as the core mass increases from $0.20\, \Msun$ to $0.46\,\Msun$. RGB tracks, as expected, move to higher $T_{\rm eff}$ as stellar mass increases, and become more luminous at higher metallicity for the same core mass.
We can see that the sequences with \texttt{\AE SOPUS\,2.0} opacities are slightly warmer in $T_{\rm eff}$, corresponding to a $\approx 3-15$ K difference. At these temperatures, the H$^-$ opacity contribution is significant, but the \cite{McLaughlin_2017} revision has little effect.


In the right panel of Figure~\ref{fig_hayashi}, we investigate the effect of the two \texttt{\AE SOPUS} versions on a TP-AGB star of given luminosity
($L=10^4\,L_{\odot}$) as the photospheric C/O increases from 0.3 to 3, as a result of a progressive carbon enrichment to the surface. This is intended to simulate the effect of the third dredge-up in a simple way.
The behavior of the effective temperature as a function of \co\ is well understood, and it reflects the abrupt change in molecular equilibrium that occurs when \co\ enters the critical range $(\co)_{\rm crit} \la \co \la 1$ (see Table~\ref{tab_sun} of this work, and Section 4.2 of \citealt{Marigo_Aringer_2009} for a detailed discussion).
In these conditions, the majority of C and O atoms are locked in the stable CO molecule, and the opacity drops dramatically, causing the effective temperature to rise. As \co\ exceeds unity and more carbon is injected into the atmosphere, the opacity caused by carbon-bearing species significantly lowers the effective temperature, a well-known property of carbon stars \citep[e.g.,][]{Marigo_02}.
When the results with \texttt{\AE SOPUS\,1.0} and \texttt{\AE SOPUS\,2.0} are compared, we see that the new opacities produces larger effective temperatures (by $\approx 10-40$ K) across the entire range of \co\ considered in the test calculations. At $\co < 1$, this should be due to a lower opacity contribution of H$_2$O, whereas at $\co  > 1$, several carbon-rich opacity sources contribute to \kR\ (e.g., CN, HCN, C$_2$, C$_3$, C$_2$H$_2$). We notice that the  discrepancies reach a minimum  at $\co \approx 1$, where molecular absorption is greatly diminished. 
\begin{figure}[h!]
    \centering
    \includegraphics[width=0.48\textwidth]{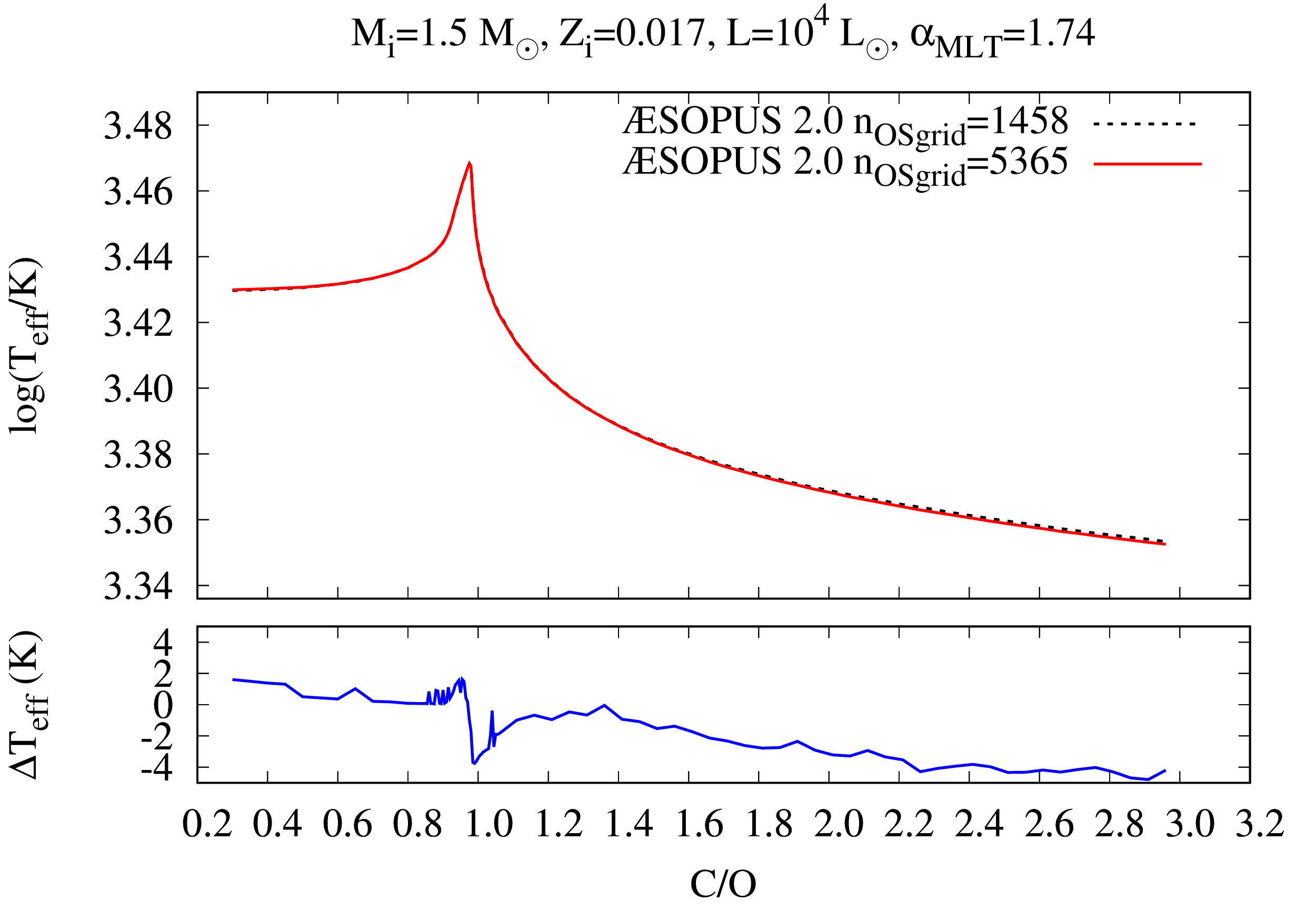}
    \caption{Effective temperature predicted by the integration of a red giant's model atmosphere. The test is designed to assess the sensitivity to the frequency grid size used to compute \kR.  {\em Top panel}:  The effective temperature of a TP-AGB star with $L=10^4\, \Lsun$ and increasing photospheric C/O, using $n_{\rm OSgrid}=5365$ (solid line) or $n_{\rm OSgrid}=1458$ (dotted line). 
    {\em Bottom panel}: Difference in effective temperature between atmospheric integrations using the two frequency grids.
    Both frequency distributions are extracted following the  \cite{Helling_Jorgensen_98} opacity sampling optimization prescriptions.  \label{fig_tpagb_grid}}
\end{figure}

In Section~\ref{ssec_molabs}, we demonstrated that reducing the number of frequency points results in small opacity differences, in the range of $0.02-0.03$ dex in $\log(\kR)$, provided the sampling distribution meets the energy requirements of \cite{Helling_Jorgensen_98} optimization scheme (see Figure~\ref{fig_osmaps}). What effect do these opacity differences have on stellar models? To answer this question, we revisited the TP-AGB star test at increasing \co\ discussed above, as shown in Figure~\ref{fig_hayashi} (right plot). This test is particularly appropriate because it covers a relevant range of effective temperatures where molecular absorption is important, as well as a wide range of chemistry configurations.

We ran two sets of  models with \texttt{\AE SOPUS\,2.0} opacities, one with $n_{\rm OSgrid}=5365$ points,  and the other with $n_{\rm OSgrid}=1458$ points. The effective temperatures derived from complete model atmospheres are compared in Figure~\ref{fig_tpagb_grid}. We can see that the differences are tiny, ranging between $-4$ K and $2$ K across the entire \co\ range. This simple experiment reassures us that even our smallest grid size preserves a high level of precision in the integration of giant stars' external layers. 
Finally, we warn the reader that a thorough investigation of the impact of the new \texttt{\AE SOPUS\,2.0} opacities  requires extensive evolutionary calculations, which will be addressed in subsequent studies.

\section{Pre-computed Opacity Tables}
\label{sec_opactables}
Using \texttt{\AE SOPUS\,2.0}, we generated a large grid of scaled-solar Rosseland mean opacity tables for a variety of initial metallicity values (from $Z=0$ to $Z=0.5$) and underlying solar mixture options. 
All opacity tables span the temperature range $3.2 \le \log(T) \le 4.5 $, and the $R$ range $-1.0 \le \log(R) \le 8.0$.
Likewise the \texttt{OPAL} opacity format \citep{OPAL_96}, for each metallicity we consider 10 potential hydrogen abundance values ($X=0, 0.1, 0.2, 0.35, 0.5, 0.7, 0.8, 0.9, 0.95, 1$), but when necessary, we reduce the number of $X$ nodes to comply with the condition that $X$ cannot exceed $1-Z$. In fact, at each metallicity, $X=1-Z$ represents always the maximum hydrogen value of the node sequence.
The Rosseland mean opacity tables are available via the repository at \url{http://stev.oapd.inaf.it/aesopus_2.0/tables}.

\section{Arbitrary Chemical Mixtures: The Need for a Web-Interface}
\label{sec_web}
\texttt{\AE SOPUS} can easily generate opacity tables for arbitrary chemical abundance distributions, such as those with varying CNO abundances, suitable for evolutionary models of red and asymptotic giant branch stars and massive rotating stars; various degrees of enhancement in $\alpha$-elements; C-N, O-Na, and Mg-Al abundance anti-correlations, which are required to properly describe the properties of stars in Galactic globular clusters; extremely metal-poor or zero-metallicity mixtures suitable for studies of gas opacity in primordial conditions, to name a few. Several applications were discussed in detail in the original paper \citep{Marigo_Aringer_2009}, and will not be repeated here.

Because the era of high-resolution and large spectroscopic surveys (e.g., \citealt[][The Gaia-ESO Large Public Spectroscopic Survey]{Randich_etal_13}; \citealt[][LAMOST]{Zhao_etal_12}; \citealt[][GALAH]{DeSilva_etal_15}; \citealt[][APOGEE]{APOGEE_17}; see also \citealt{Jofre_etal_19} for a comprehensive review) has been revealing a wide variety of abundance patterns in stars, creating archives of  opacity tables for any scenario makes little sense. A profitable way to deal with such abundance data richness is to provide a public web-interface where users can personalize their opacity query.

In this perspective and to greatly increase the availability of low-temperature opacities, \cite{Marigo_Aringer_2009} created an interactive web-interface (\url{http://stev.oapd.inaf.it/aesopus}) that allows users to run \texttt{\AE SOPUS\,1.0} based on their specific needs simply by entering the input parameters ($T-R$ grid, reference solar mixture, metallicity, abundance of each chemical species) on the web mask.
 The interface has now been updated to include the major revision introduced in \texttt{\AE SOPUS\,2.0},  while maintaining the high level of flexibility and quick computational performance that distinguishes our public tool. It is accessible via the URL \url{http://stev.oapd.inaf.it/aesopus_2.0}.
 The previous web-interface, corresponding to \texttt{\AE SOPUS\,1.0}, is still available under the URL \url{http://stev.oapd.inaf.it/aesopus_1.0}.

\section{Concluding Remarks}
\label{sec_conclu}
Updated \texttt{\AE SOPUS} low-temperature opacities have been computed for various solar mixtures and made publicly available for primary use in stellar models. The tables can be obtained through a static repository at  \url{http://stev.oapd.inaf.it/aesopus_2.0/tables}.
The major changes are improved input physics and numerical procedures to speed up computational effort,  while maintaining  high accuracy. Among the updates are recommended line lists for 80 absorbing molecular species from the \texttt{ExoMol} and \texttt{HITRAN} databases, new data for H$^-$ photodetachment bound-free absorption, and revised collision-induced absorption. The most recent solar mixtures from \citet{Asplund_etal_21} and \citet{Magg_etal_22} are added. 
The \texttt{\AE SOPUS} web-interface has been renovated to integrate all the changes introduced in this work. It is available through the URL \url{http://stev.oapd.inaf.it/aesopus_2.0}. User feedback is encouraged.

This significant update in the \texttt{\AE SOPUS} code and related deliverables is only the first step in a major revision and expansion of our tools for dealing with opacity tables. Following works will address opacities in the high-temperature regime, opacities of heavy elements such as Lanthanides and Actinides, systematic update of partition functions, novel interpolation schemes.  Furthermore, any opacity revisions  will be incorporated and tested in the \texttt{PARSEC} \citep{Bressan_etal_12,costa19} and \texttt{COLIBRI} \citep{colibri} stellar evolutionary codes.

Finally, we conclude with a few thoughts on the needs we believe are critical to improving the opacities for the stellar community.
The availability of accurate and comprehensive energy levels, line positions, oscillator strengths, and cross sections for  significant absorbing species (atoms, ions, anions, molecules and transitory dipoles produced in collisions) is required for robust and reliable integration of Rosseland mean opacities.
There has been a lot of work done in recent years to build extensive molecular line lists, such as the coordinated project \texttt{ExoMol}, which is primarily designed for exoplanets. \texttt{ExoMol} also provides a suite of user-friendly tools for computing partition functions and opacity cross sections. The included energy levels are usually complete enough to cover the typical temperatures of the stellar atmospheric layers where molecules can form.

For cool stars with $\co<1$, the situation with the line lists is quite favorable. For example, we can now rely on previously unavailable data for molecules such as AlH, NaH, and CaOH, which are essential for modeling the atmospheres of M dwarfs.
Still, further effort needs to be done to improve stellar opacities. 
Pressure broadening should be taken into account for both M and brown dwarfs. The modeling is complex, and it is partly hampered by the poor knowledge of the broadening parameters for major collision broadeners, such as H$_2$ and He, at relatively high temperatures.

Transition metal bearing diatomic molecules are important opacity sources at near infrared and visible wavelengths. We have line lists for several species, including TiO, VO, FeH, ScH, TiH, CrH, NiH, ZrO, and YO, but there are many other candidates that lack data (e.g., MnH, FeO, TiZr, ZrV).
The chemistry of carbon stars with $\co>1$ is more complex.
For some molecules, such as C$_3$, which is an important opacity source in cool carbon stars with a high \co\ ratio, we still rely on old line lists \citep{Jorgensen_etal_89}, that deserve to be improved. In fact, it is well known that the existing computed opacities do not accurately reproduce the observed spectral features of this species \citep{Aringer_etal_19}. 
Furthermore, there could be other significant opacity sources in carbon stars for which we have no data at all (e.g.\ C$_2$H).
The cross sections of some bound-free and free-free processes, such as those of anions of atoms and molecules, are estimated from early studies carried out many decades ago (see Table~\ref{tab_opacsource}). A modern revision would undoubtedly be beneficial. 

\begin{acknowledgments}
We acknowledge support from Padova University through the research project PRD 2021. 
Also special thanks to Jonathan Tennyson for helpful advice on the use of the \texttt{ExoMol} database, Maria Bergemann for useful discussion on solar abundances, Sophie Van Eck and Bertrand Plez for providing us with the ZrO line list.
\end{acknowledgments}

\software{\texttt{\AE SOPUS} \citep{Marigo_Aringer_2009},  
    \texttt{EXOCROSS} \citep{EXOCROSS_2018}, \texttt{COLIBRI} \citep{colibri}
          }



\bibliography{aesopus2.bib}{}
\bibliographystyle{aasjournal}



\end{document}